\documentclass[a4paper,onecolumn,accepted=2021-07-26]{quantumarticle}
\pdfoutput=1

\usepackage{physics}
\usepackage{amsmath,amssymb}
\usepackage{mathtools}
\usepackage{graphicx}
\usepackage{dcolumn}
\usepackage{bm}
\usepackage{color,soul}
\usepackage{array}
\usepackage{amsthm}
\usepackage{subcaption}
\usepackage{algorithm}
\usepackage{algpseudocode}
\usepackage{wrapfig}
\usepackage{qcircuit}
\usepackage[colorlinks,linkcolor=blue,anchorcolor=red,citecolor=red]{hyperref}
\usepackage[capitalize]{cleveref}
\usepackage[numbers,sort&compress]{natbib}
\usepackage[skins]{tcolorbox}

\newtheorem{theorem}{Theorem}

\algnewcommand\algorithmicswitch{\textbf{switch}}
\algnewcommand\algorithmiccase{\textbf{case}}
\algnewcommand\algorithmicassert{\texttt{assert}}
\algdef{SE}[SWITCH]{Switch}{EndSwitch}[1]{\algorithmicswitch\ #1\ \algorithmicdo}{\algorithmicend\ \algorithmicswitch}%
\algdef{SE}[CASE]{Case}{EndCase}[1]{\algorithmiccase\ #1}{\algorithmicend\ \algorithmiccase}%
\algtext*{EndSwitch}%
\algtext*{EndCase}%
\algnewcommand{\IIf}[1]{\State\algorithmicif\ #1\ \algorithmicthen}
\algnewcommand{\EndIIf}{\unskip\ \algorithmicend\ \algorithmicif}

\begin{document}
	
	\title{Quantum walk-based portfolio optimisation}
	
	\author{N. Slate}
	\affiliation{Department of Physics, The University of Western Australia, Perth WA 6009, Australia}
	\author{E. Matwiejew}
	\affiliation{Department of Physics, The University of Western Australia, Perth WA 6009, Australia}
	\author{S. Marsh}
 	\email{samuel.marsh@research.uwa.edu.au}
	\affiliation{Department of Physics, The University of Western Australia, Perth WA 6009, Australia}
	\author{J. B. Wang}
	\email{jingbo.wang@uwa.edu.au}
	\affiliation{Department of Physics, The University of Western Australia, Perth WA 6009, Australia}

	\begin{abstract}
	This paper proposes a highly efficient quantum algorithm for portfolio optimisation targeted at near-term noisy intermediate-scale quantum computers. Recent work by Hodson et al. (2019) explored potential application of hybrid quantum-classical algorithms to the problem of financial portfolio rebalancing. In particular, they deal with the portfolio optimisation problem using the Quantum Approximate Optimisation Algorithm and the Quantum Alternating Operator Ansatz. In this paper, we demonstrate substantially better performance using a newly developed Quantum Walk Optimisation Algorithm in finding high-quality solutions to the portfolio optimisation problem.
	\end{abstract}
	
	\maketitle
	
	\section{Introduction}

    Quantum computers are powerful devices that utilise intrinsic properties of quantum mechanics, such as superposition and entanglement, to provide substantial speedups for solving computationally hard problems \cite{Shor_1997, nielsen2010quantum}. A recent influx of interest and technological advancements in this field have lead to the discussion of practical applications especially in the Noisy Intermediate-Scale Quantum (NISQ) era~\cite{Preskill_2018}. This includes solving difficult financial problems \cite{Rebentrost2018a, Rebentrost2018b, Or_s_2019, Woerner2019}; one such  problem is portfolio optimisation and periodic re-balancing~\cite{CBA}. When considering a set of $n$ assets there are $3^n$ different portfolio combinations when we consider three different discrete asset positions:
    \begin{enumerate}
\item Long position: the buying of an asset such as a stock, commodity or currency with the expectation that it will rise in value;
\item Short position: the selling of an asset with the expectation that it will drop in value;
\item No position: neither a long nor short position is taken.
    \end{enumerate}

    In this work, we take the mean-variance Markowitz model \cite{mark, modernalgo} as the basis for portfolio optimisation, which is fundamental to modern portfolio theory. Although this model was developed in the 1950s, its simplicity, relative accuracy and relevance persists as an area of exploration for the quantum computing community~\cite{CBA}.
    
    Financial portfolio optimisation and the Markowitz model have been shown, when discrete asset constraints are involved, to fall into the category of NP-hard combinatorial optimisation problems \cite{MansiniRenata1999Haft,Coleman2006}. Thus, the portfolio optimisation problem provides a real-world model to investigate quantum speedups through quantum approximate optimisation algorithms. Of interest are implementations appropriate for near-term noisy intermediate-scale quantum (NISQ) computers, which have become increasingly available in the cloud, and are fast approaching sufficient scale and fidelity \cite{Preskill_2018}.  
    
    The problem of portfolio optimisation and rebalancing using the Markowitz model with discrete asset constraints has been previously evaluated using the Quantum Approximate Optimisation Algorithm and Quantum Alternating Operator Ansatz \cite{CBA}. Collectively known as QAOA, the algorithms were developed by \cite{cook2019quantum} and \cite{farhi2016quantum} to solve combinatorial optimisation problems. The latter algorithm generalises the original QAOA to incorporate optimisation constraints. Since we aim to compare these algorithms in this paper, we distinguish them as QAOA and QAOAz respectively. These algorithms are known as hybrid quantum-classical algorithms, as they utilise both quantum computing advantages and classical optimisation in order to minimise a given objective function. QAOA can be thought of as a Trotterised approximation to the Quantum Adiabatic Algotithm (QAA) \cite{farhi2016quantum}, where the system is evolved into the ground state of some operator that encodes the problem solution. As a heuristic or approximation algorithm, QAOA accepts high-quality solutions when the optimal solution is not found.
   
   Specifically, QAOA-based algorithms utilise an alternating state evolution consisting of solution-quality-dependant phase shifts and a mixing of probability amplitude across a state-space of possible solutions. Using a hybrid quantum-classical variational scheme, the expectation value of an operator encoding the objective function of an associated scalar optimisation problem is minimised; such that the probability of measuring the system in a state corresponding to a high quality solution is amplified. 
   
   This paper evaluates a further development in QAOA schema known as the Quantum Walk Optimisation Algorithm (QWOA) \cite{marshqwoa}. QWOA utilises an efficient indexing algorithm in conjunction with a generalisation of the QAOA mixing operator to a continuous-time quantum walk over a circulant graph that connects all feasible solutions. Our earlier work indicated that QWOA offers significant advantages over pre-existing methods through a reduction in the search space and an unbiased encoding of optimisation constraints \cite{marshqwoa}. In this paper we provide numerical evidence for the efficacy of QWOA through its application to portfolio optimisation. 
   
The paper is organised as follows. In Sec. IIA, we introduce the Markowitz model for portfolio optimisation, along with its quantum encoding for approximate optimisation. In Sec. IIB, we compute the size of the search space for QAOAz and QWOA. This is followed by a detailed description and circuit comparison of the three quantum approximate optimisations under consideration. Finally, in Sec. III and Sec. IV, we present numerical results and analysis.
    
    \section{Portfolio Optimisation Problem Formulation}
    
    The formalisation and models used in this paper are based upon the work done by \citet{CBA} as to provide a basis for comparison between QAOA, QAOAz, and QWOA.
    
    \subsection{The Markowitz model}
    The discrete mean-variance Markowitz model can be expressed through minimisation of the objective function
    \begin{equation}
        c(z) = \lambda\sum_{i,j=1}^{n}\sigma_{ij}z_iz_j - (1 - \lambda)\sum_{i=1}^{n}r_iz_i \,
        \label{eq:markowitz}
    \end{equation}
    which we subject to the optimisation constraint
    \begin{equation} \label{eq:constraint}
    \sum_{i=1}^{n}z_i = A \, .
    \end{equation}
    {In the above formulation, $z=z_1 \ldots z_n$ encodes a particular choice of portfolio from $n$ assets, where for each asset $i$ we have $z_i \in \{1,-1,0\}$ representing the choice of a long position, a short position, or no position. Associated with each asset is an expected return $r_i$, and the correlation between two assets $i$ and $j$ is given by the covariance $\sigma_{ij}$. These values are derived from historical data.
    The risk aversion parameter $\lambda$, taking a value between $0$ and $1$, reflects the balance between returns and risk. As this risk parameter approaches $0$, the optimal portfolio is based purely on obtaining maximum returns. As $\lambda$ approaches 1, the only consideration becomes minimisation of risk (for example, leading to a preference for government bonds over real estate). The value $A$ defines the net total of discrete lots to be invested for the portfolio.
    Note that in the definition of $A$, the signed quantity $z_i$ is summed rather than the absolute value $\abs{z_i}$, which is a formulation generally used in portfolio \textit{rebalancing} to treat the relative net position with respect to an existing portfolio {\cite{CBA}}.
    
    The mean-variance formulation is standard and well-studied in finance, both in the continuous and discrete domain {\cite{Steinbach2001,Anagnostopoulos2011}}. The model formalises the idea of portfolio diversification, where an investor can reduce risk by owning a number of assets having low correlations on returns. The application of discrete variables commonly applies to situations where the class of assets is traded in discrete quantities (e.g. real estate), and when the cost of a single asset is substantial (bonds often meet this criteria). It additionally is applicable where one wishes to subdivide a fixed budget equally amongst the assets in the portfolio.}
    
    In order to `quantise' the formulation and encode each portfolio using a quantum register, we require two qubits per asset as shown in \cref{tbl:qubitportfolio}, representing the three possible position choices for each asset.
    
    \begin{table}[h!]
    \begin{center}
    \begin{tabular}{ | c | c | c | c | } 
    \hline
     `Short' qubit & `Long' qubit & $z_i$ value & $\ket{z}$ encoding \\ 
    \hline
    \hline
    $\ket{0}$ & $\ket{0}$ & $0$ & $\ket{00} $\\ 
    $\ket{0}$ & $\ket{1}$ & $1$ & $\ket{01}$ \\ 
    $\ket{1}$ & $\ket{0}$ & $-1$ & $\ket{10}$ \\
    $\ket{1}$ & $\ket{1}$ & $0$ & $\ket{11}$ \\
    \hline
    \end{tabular}
    \end{center}
    \caption{Qubit encoding of the possible asset positions.}
    \label{tbl:qubitportfolio}
    \end{table}
    
    \subsection{Counting valid portfolios}
    
    For the QWOA, it is a necessary condition \cite{marshqwoa} to identify the cardinality of the solution set for any given parameters $(n, A)$. It is also useful to see the difference in the size of the constrained search subspace to compare between QAOA, QAOAz and QWOA.

    As there are $n$ assets under consideration, and each asset uses two qubits to encode its position, the Hilbert space is of size $N=2^{2n}$, which is the ordinary QAOA search space. However, only a subset of these states correspond to valid portfolio configurations that satisfy the constraint given in \cref{eq:constraint}. Furthermore, portfolio configurations that include a stock with both position qubits set to $\ket{1}$ are degenerate -- the $\ket{11}$ configuration is interpreted as equivalent to the $\ket{00}$ configuration, i.e. no position. This means there can be a large amount of degeneracy amongst the `valid' states.
    
\begin{figure}[hb]
        \centering
        \includegraphics[width=0.8\linewidth]{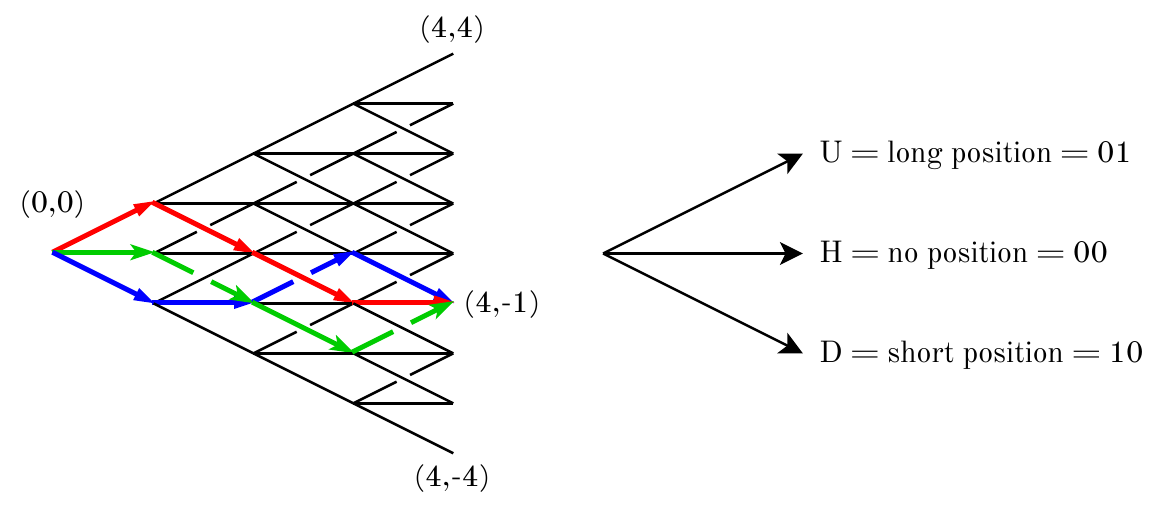}
        \caption{An illustrative lattice representation of three example portfolios which contain four assets and require one more short position than long position. The red, blue and green paths represent the portfolios 01101000, 10000110 and 00101001 respectively. There are a total of $\mathcal{M}(4, -1)=16$ paths that reach this endpoint.}
        \label{fig:lattice}
    \end{figure}

    We aim to count the number of valid and non-degenerate states, which we will later see corresponds to the size of the search subspace for QWOA. First, we note that the number of valid nondegenerate $(n,A)$-portfolios has an exact correspondence with the number of lattice paths from $(0, 0)$ to $(n, A)$ with steps taken from the set $\{U=(1,1),D=(1,-1),H=(1,0)\}$. An example of this correspondence is given in \cref{fig:lattice}. Choosing a long position on a stock is equivalent to stepping diagonally up, while choosing a short position is equivalent to stepping diagonally down. Choosing no position on a stock takes a step horizontally. To satisfy the investment constraint of having $A$ more long positions than short positions, the path must end at $(n, A)$.

\begin{figure}[h!]
        \centering
        \includegraphics[width=0.6\linewidth]{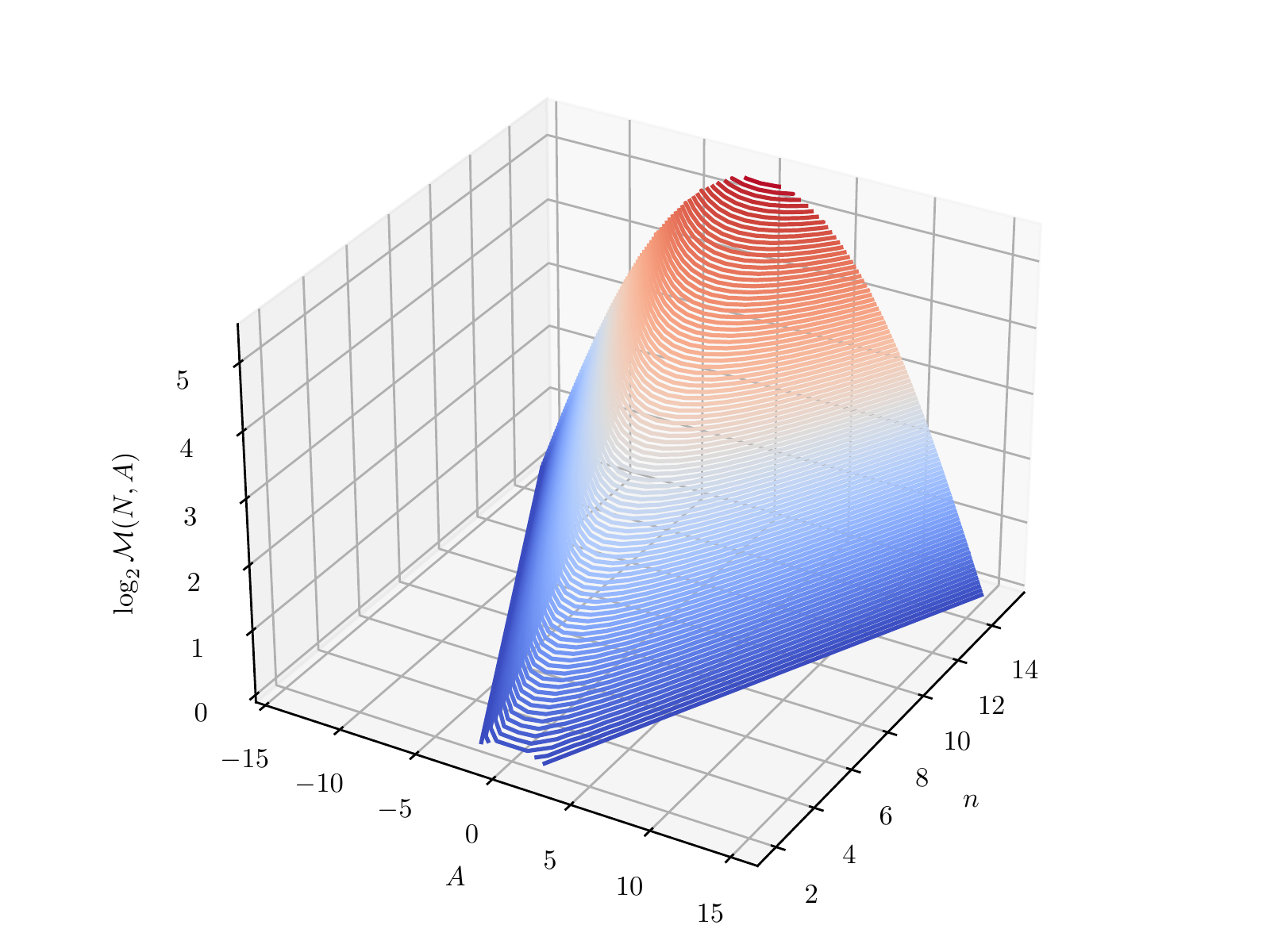}
        \caption{Log-number of valid non-degenerate portfolios for values of $n$ and $A$.}
        \label{fig:numsols}
    \end{figure}

    \begin{theorem}
        The number of such lattice paths, plotted in \cref{fig:numsols}, is given by
        \begin{equation}
            \mathcal{M}(n,A)=\sum\limits_{j=0}^n \binom{n}{j} \binom{n-j}{\frac{1}{2}(n+A-j)}
            \label{eq:numvdsolutions}
        \end{equation}
        where the second binomial coefficient is set to $0$ if its bottom parameter is not an integer.
    \end{theorem}
    
    \begin{proof}
        We follow the techniques used in \cite{Bona2015}. In particular, the lattice path is similar to a so-called \textit{Motzkin path}, for which the number of paths is given by Eq. (10.43) in \cite{Bona2015}. Suppose that there are $j$ $H$-moves in a given path from $(0,0)$ to $(n,A)$. If these are removed, we are left with a path from $(0,0)$ to $(n-j,A)$ using only diagonal moves. Performing the coordinate mapping $(x,y) \mapsto (\frac{1}{2}(x+y),\frac{1}{2}(x-y))$ transforms the problem into finding a path to $(\frac{1}{2}(n+A-j),\frac{1}{2}(n-A-j))$ using standard basis steps $(1,0)$ and $(0,1)$. The number of paths to this endpoint with standard basis steps, using Eq. (10.3) in \cite{Bona2015}, is
        \begin{equation}
            \binom{n-j}{\frac{1}{2}(n+A-j)} \, ,
        \end{equation}
        where the binomial coefficient is 0, if the bottom component is fractional.
        There are $\binom{n}{j}$ ways to re-insert the originally removed $H$-moves, and after summing over $j$ we arrive at the desired result.
    \end{proof}
    
   The search space of QAOAz is larger than $\mathcal{M}(N, A)$, since we will later see that it incorporates valid but also degenerate solutions. The number of all valid portfolio encodings, including degeneracies, can be expressed as 
    \begin{equation} \label{eq:QAOAz_size}
        \mathcal{V}(n,A)=\sum\limits_{j=0}^n 2^j \binom{n}{j} \binom{n-j}{\frac{1}{2}(n+A-j)} = \binom{2n}{n+A} \, ,
    \end{equation}
    since if there are $j$ $H$-moves the corresponding $j$ stocks may each be encoded as $\ket{00}$ or $\ket{11}$. More simply, the second expression is derived from choosing $(n+A)$ bits to flip in the $2n$-qubit state encoding. In the following section it will be shown that this is the size of the search space for QAOAz.
    
    \subsection{Quantum Approximate Optimisation Algorithms}\label{3type}
    
   The Quantum Approximate Optimisation Algorithms evolve a given initial state as
    \begin{equation}         \label{QAOAevolution}
    \ket{\psi_f}=\prod_{i=1}^{p}U_\text{mix}(t_i)U_C(\gamma_i)\ket{\psi_0} \, .
    \end{equation}
    {Here, $U_C(\gamma) \ket{z} = e^{-i \gamma c(z)} \ket{z}$ for any portfolio qubit encoding $\ket{z}$ as per {\cref{tbl:qubitportfolio}}, applying a relative phase shift to the solution proportional to the variational parameter $\gamma$ and to the value of the portfolio with respect to the objective function. This operator, as defined based on the objective function $c(z)$ given in {\cref{eq:markowitz}}, is the same for all methods discussed in this section. We also define the corresponding Hamiltonian $C$ such that $U_C(\gamma) = e^{-i \gamma C}$, which encodes the values of the objective function $c(z)$ as its diagonal elements.} The operator $U_\text{mix}$ is the mixing operator that differs depending on the choice of connectivity between solution states. In this section, we contrast three different algorithms that have differing $U_\text{mix}$ operators and associated initial states $\ket{\psi_0}$. The integer parameter $p$ is the number of QAOA iterations, with increased $p$ providing improved solution quality at the cost of increased circuit depth and $(\vec{t}, \vec{\gamma})$ are circuit parameters which are classically optimised in order to minimise the expectation value $\bra{\psi_f}C\ket{\psi_f}$. In this framework, a lower expectation value corresponds to a greater probability of measuring a portfolio with an optimal, or near optimal, balancing of predicted future returns and risk.

    \subsubsection{Quantum Approximate Optimisation Algorithm (QAOA)}
    
    For the original QAOA scheme, the mixing operator $U_\text{mix}$ is defined as $U_\text{QAOA}(t) = e^{-i t \sum_i X^{(i)}}$. 
    One method of encoding the investment constraint is to incorporate it within the objective function. This so-called `soft constraint' occurs as the addition of a penalty function to \cref{eq:markowitz},
    \begin{equation}
        \zeta(z)=\epsilon(A-\sum_{i=1}^{n}{z_i})^2
    \end{equation}
    \noindent where $\epsilon > \max(c(z))-\min(c(z))$.
     Thus for QAOA applied to portfolio optimisation, $C\ket{z} = (c(z) + \zeta(z))\ket{z}$. Final states with minimal expectation value correspond to `good' solutions to the given optimisation problem.
     The penalty function penalises portfolios with a net long position of more than or less than $A$ assets. With the use of the $\epsilon$ inequality, this penalty function ensures that the minimum objective function value corresponds to a state which satisfies the constraint. The soft constraint method enables the QAOA algorithm to optimise given the constraint, but there are states in which the constraint is not satisfied which are still being considered by the algorithm.
    
    The initial state for the soft constraint method is simply an equal superposition across all states,
    \begin{equation}
        \ket{\psi_{0}}=\frac{1}{\sqrt{2^{2n}}}\sum_{i=0}^{N-1}\ket{i} \, .
        \label{eq:softstart}
    \end{equation}
    
    \subsubsection{Quantum Alternating Operator Ansatz (QAOAz)} \label{sec:QAOAz}
    
    This extension of the QAOA provides a means of constraining the optimisation process to the subspace of valid solutions by modification of the mixing operator. The phase unitary, $U_C(\gamma)$, maintains the same form for all algorithms discussed in this paper. However, the mixing operator is now replaced by $U_{\text{QAOAz}}$, which is an approximation to the time evolution of the `dual parity ring' Hamiltonian as given in \cite{CBA,Hadfield_2019}, 
	\begin{equation}
	    \label{eq:dualtransversefield}
	    {H_\text{ring} = \sum\limits_{a=1}^{2n} \left( X^{(a)}X^{(a+2)} + Y^{(a)}Y^{(a+2)} \right) \, ,}
	\end{equation}
	where all addition is modulo $n$.
	This Hamiltonian preserves the Hamming weight of both the short and long qubits independently, and consequently meets the $A$-constraint.
    \citet{Hadfield_2019} provide a method for approximating the time evolution of the dual parity ring mixer Hamiltonian by decomposing it into three non-commuting unitaries. The QWOAz mixer for portfolio optimisation is thus defined as
    \begin{equation} \label{eq:qaoaz}
        U_\text{QAOAz}(t)=U_{\text{last}}(t)U_{\text{even}}(t)U_{\text{odd}}(t) \approx e^{-i t H_\text{ring}}
    \end{equation}
    where
    \begin{align}
        U_{\text{odd}}(t) &= \prod_{\substack{a \ \text{odd}, \\ a\neq n}} e^{-i t \left( X^{(2a+1)}X^{(2(a+2))}+Y^{(2a+1)}Y^{(2(a+2))}\right)}e^{-i t \left( X^{(2(a+1))}X^{(2(a+2))}+Y^{(2(a+1))}Y^{(2(a+2))}\right)} \, ,\\
        U_{\text{even}}(t) &= \prod_{a \ \text{even}} e^{-i t \left( X^{(2a+1)}X^{(2(a+2))}+Y^{(2a+1)}Y^{(2(a+2))}\right)}e^{-i t \left( X^{(2(a+1))}X^{(2(a+2))}+Y^{(2(a+1))}Y^{(2(a+2))}\right)} \, ,\\
        U_{\text{last}}(t) &= \begin{dcases}
            e^{-i t (X^{(2n)} X^{(1)} + Y^{(2n)} Y^{(1)})}e^{-i t (X^{(2n-1)} X^{(2)} + Y^{(2n-1)} Y^{(2)})}, & n \ \text{odd}, \\
            I, & n \ \text{even}. \\
        \end{dcases}
    \end{align}
    \noindent where again all addition is modulo $n$. In the above expressions, the first exponential acts independently on the short position qubits, whilst the second acts on the long position qubits. When given an initial state satisfying \cref{eq:constraint}, the action of the dual parity ring mixer ensures that QAOAz has non-zero amplitude only in solution states satisfying the net investment constraint.
    
  The mixing operator $U_\text{QAOAz}$ creates a structure known as parity bands, which are a result of the preservation of both the long and short qubits independently. Given there are more than one way to satisfy the investment constraint (e.g. for $n=5$ and $A=4$, two valid portfolios are 4 long positions and 1 no position, or 5 long positions and 1 short position), the parity bands are disconnected. Consequently, there is no ability to transfer amplitude between parity bands through this mixing operator. 
  We must thus ensure the initial state is in a superposition across all possible valid parity bands. 
  Given $n$ assets with a net portfolio position of $A$, a simple initial state given in \cite{CBA} that meets this criteria is
  
    \begin{equation}
        \ket{\psi_0} = \frac{1}{2^{(n-A)/2}}\ket{01}^{\otimes A}\otimes (\ket{00}+\ket{11})^{\otimes(n-A)} \, ,
    \end{equation}
    which represents an equal superposition over all the (degenerate) portfolios having exactly $A$ more long positions than short positions and $(n-A)$ no-positions.
    This initial state is efficient to prepare, but at the cost of bias -- the probability amplitude is binomially distributed across parity bands \cite{CBA}.

    \subsubsection{Quantum Walk Optimisation Algorithm (QWOA)}
    \label{sec:finance-qwoa-method}
    
    QWOA generalises the original QAOA mixer as a continuous-time quantum walk (CTQW) over the subspace of valid solutions~\cite{marshqwoa}. QWOA again evolves the initial state as given by \cref{QAOAevolution}, but with a new quantum walk mixer $U_\text{QWOA}$. The quantum analogy to the classical random walk, a CTQW is the evolution of a quantum system under a Hamiltonian defined by a graph adjacency matrix \cite{CTQWFarhi, qwbook2014}. The advantage of QWOA for the portfolio optimisation problem lies in its  flexibility in `connecting' only the solutions in the valid subspace, the ability to eliminate degenerate portfolio states (thus significantly reducing the search space), and complete global symmetry amongst valid solutions (eliminating bias of one valid solution over another due to mixing asymmetry). 
   
    In order to implement a QWOA mixer on a desired subspace of solutions $\mathcal{S}$, we first design an efficiently computable bijection $\mathcal{S} \xrightarrow{\text{id}} \{0, 1, \ldots, \abs{\mathcal{S}}-1\}$ \cite{marshqwoa}, as illustrated in \cref{fig:index}. In the case of portfolio optimisation, we need a classical algorithm $\text{id}_{n,A}(x)$ to index the valid and canonical (non-degenerate) portfolio encodings $x$. In the following, we provide such an algorithm to compute $\text{id}_{n,A}(x)$ for any given valid portfolio $x$ and $\text{id}_{n,A}^{-1}(j)$ for any given index $j$.
    
    \begin{table}[h!]
      \[ \left[
        \begin{array}{cc}
         x & \text{id}(x) \\ \hline
         01010000_2 & 0 \\
         01000100_2 & 1 \\
         00010100_2 & 2 \\
         01000001_2 & 3 \\
         00010001_2 & 4 \\
         00000101_2 & 5 \\
         10010101_2 & 6 \\
         01100101_2 & 7 \\
         01011001_2 & 8 \\
         01010110_2 & 9 \\
        \end{array}
        \right] \]
        \centering
      \caption{Indexing function for $n=4,A=2$, a bijection between the portfolio binary encodings and the integers.}
    \label{fig:index}
    \end{table}
    
    We use a simple recursion relation to index, based on the counting function in \cref{eq:numvdsolutions}. Observe that $\mathcal{M}(n, A)=\mathcal{M}(n-1, A)+\mathcal{M}(n-1, A-1)+\mathcal{M}(n, A+1)$. Using the lattice analogy, the number of paths reaching $(n, A)$ is the sum of the number of paths one step from $(n, A)$. This inspires a recursive ranking algorithm as per \cref{alg:index}. The un-indexing algorithm works similarly, mapping an integer index to a binary portfolio representation as per \cref{alg:unindex}.
    
    \begin{algorithm}[H]
		\caption{$\text{id}(n, A, x)$}
		\label{alg:index}
		\begin{algorithmic}
		    \IIf{$x=0$} \Return 0 \EndIIf
			\State $x_f \gets x \mathbin{\&} 11_2$ \Comment{last two bits of $x$}
			\State $x' \gets x \gg 2$ \Comment{remove last two bits}
			\Switch{$x_f$}
                \Case{$00_2$:} $r \gets \text{id}(n-1, A, x')$ \Comment{$H$-move} \EndCase 
                \Case{$01_2$:} $r \gets \mathcal{M}(n-1, A) + \text{id}(n-1, A-1, x')$ \Comment{$U$-move} \EndCase
                \Case{$10_2$:} $r \gets \mathcal{M}(n-1, A) + \mathcal{M}(n-1, A-1) + \text{id}(n-1, A+1, x')$ \Comment{$D$-move} \EndCase
            \EndSwitch
        \State \Return r
		\end{algorithmic}
	\end{algorithm}
	\begin{algorithm}[H]
		\caption{$\text{id}^{-1}(n, A, j, x=0)$}
		\label{alg:unindex}
		\begin{algorithmic}
		    \IIf{$n=0$} \Return $x$ \EndIIf
			\If{$j<\mathcal{M}(n-1, A)$} \Comment{$H$-move}
			    \State $r \gets \text{id}^{-1}(n-1, A, j, x) \ll 2$ 
			\ElsIf{$j<\mathcal{M}(n-1, A)+\mathcal{M}(n-1, A-1)$} \Comment{$U$-move}
			    \State $j' \gets j - \mathcal{M}(n-1, A)$
			    \State $r \gets 01_2 \mathbin{|} (\text{id}^{-1}(n-1, A-1, j', x) \ll 2)$
			\Else \Comment{$D$-move}
			    \State $j' \gets j - (\mathcal{M}(n-1, A)+\mathcal{M}(n-1, A-1))$
			    \State $r \gets   10_2 \mathbin{|} (\text{id}^{-1}(n-1, A+1, j', x) \ll 2)$
			\EndIf
        \State \Return r
		\end{algorithmic}
	\end{algorithm}
	
\begin{figure*}[h!]
    \begin{subfigure}{\textwidth}
      \centering
      \[
         \Qcircuit @C=1em @R=0.7em {
            \lstick{\ket{A}} & {/} \qw & \multigate{1}{\mathcal{M}^{j}_0} & \qw & \qw & \qw & \multigate{1}{(\mathcal{M}^{j}_0)^\dag} &  \push{\rule{.3em}{0.25pt}}\qw & \multigate{1}{\mathcal{M}^{j}_{-1}} & \qw & \qw & \qw & \multigate{1}{(\mathcal{M}^{j}_{-1})^\dag} & \gate{+1} & \gate{-1} & \qw & \rstick{\ket{A'}}\\
            \lstick{\ket{0}} & {/} \qw & \ghost{\mathcal{M}^{j}_0} & \ctrl{1} & \qw & \multigate{1}{-} & \ghost{(\mathcal{M}^{j}_0)^\dag} & \qw & \ghost{\mathcal{M}^{j}_{-1}} & \ctrl{1} & \qw & \multigate{1}{-} & \ghost{{(\mathcal{M}^{j}_{-1})^\dag}} & \qw & \qw & \qw & \rstick{\ket{0}}\\
            \lstick{\ket{y}} & {/} \qw & \push{\rule{0em}{1em}}\qw & \ctrl{3} & \qw & \ghost{-} & \qw & \qw & \qw & \ctrl{2} & \qw & \ghost{-} & \qw & \qw & \qw & \qw & \rstick{\ket{y'}}\\
            \lstick{\cdots} & & & & & & & && & & &  \\
            \lstick{\ket{0_j}} & \qw & \qw & \push{\rule{0em}{1em}} \qw & \qw & \qw & \qw & \qw & \qw & \gate{<} & \qw & \ctrl{-2} & \ctrl{1} & \qw & \ctrl{-4} & \qw & \rstick{\ket{x_{j}}}\\
            \lstick{\ket{0_{j+1}}} & \qw & \push{{\rule{0em}{1em}}} \qw & \gate{<} & \qw & \ctrl{-3} & \qw & \qw & \push{{\rule{0em}{1em}}} \qw & \qw & \qw & \qw & \targ & \ctrl{-5} & \qw & \qw & \rstick{\ket{x_{j+1}}}\\
            \lstick{\cdots}
            \gategroup{1}{3}{3}{15}{1em}{.}
            \gategroup{5}{4}{6}{15}{1em}{.}
        }
        \]
        \caption{}
      \label{fig:sub-first}
    \end{subfigure}
    \begin{subfigure}{\textwidth}
      \centering
        \[
         \Qcircuit @C=1em @R=0.7em {
            \lstick{\ket{A}} & {/} \qw & \multigate{1}{\tilde{U}_{n-1}} & \qw & & \multigate{1}{\tilde{U}_{1}} & \qw & \rstick{\ket{0}} &  \\
            \lstick{\ket{y}} & {/} \qw & \ghost{\tilde{U}_{n-1}} & \qw & & \ghost{\tilde{U}_{1}} & \qw & \rstick{\ket{0}}  & \\
            \lstick{\ket{0_1}} & \qw & \qw & \qw & & \multigate{1}{\tilde{U}_{1}} \qwx[-1] & \qw & \rstick{\ket{x_1}}  & \\
            \lstick{\ket{0_2}} & \qw & \qw & \qw & & \ghost{\tilde{U}_{1}} & \qw & \rstick{\ket{x_2}}  & \\
            \lstick{\cdots} &  & & & \push{\cdots} & & & \rstick{\cdots}  & &  \rstick{\ket{\text{id}^{-1}(n, A, y)}} \\
            \lstick{\ket{0_{n-1}}} & \qw & \multigate{1}{\tilde{U}_{n-1}} \qwx[-4] & \qw & & \qw & \qw & \rstick{\ket{x_{n-1}}} &  \\
            \lstick{\ket{0_{n}}} & \qw & \ghost{\tilde{U}_{n-1}} & \qw & & \qw & \qw & \rstick{\ket{x_{n}}} & \push{\rule{3em}{0em}}
            \gategroup{3}{9}{7}{9}{1em}{\}}
        }
        \]
        \caption{}
      \label{fig:sub-second}
    \end{subfigure}
    \caption{(a) The $j$th sub-circuit for QWOA portfolio un-indexing, which we label as $\tilde{U}_j$. This circuit, given the index $y$, decodes the $j$th and $(j+1)$th bits of the portfolio representation. (b) Illustration of the quantum un-indexing circuit $U_\#^\dag$ for QWOA portfolio optimisation. This performs the unitary mapping $\ket{y} \mapsto \ket{\text{id}^{-1}(n, A, y)}$.}
    \label{fig:qcircuit}
    \end{figure*}
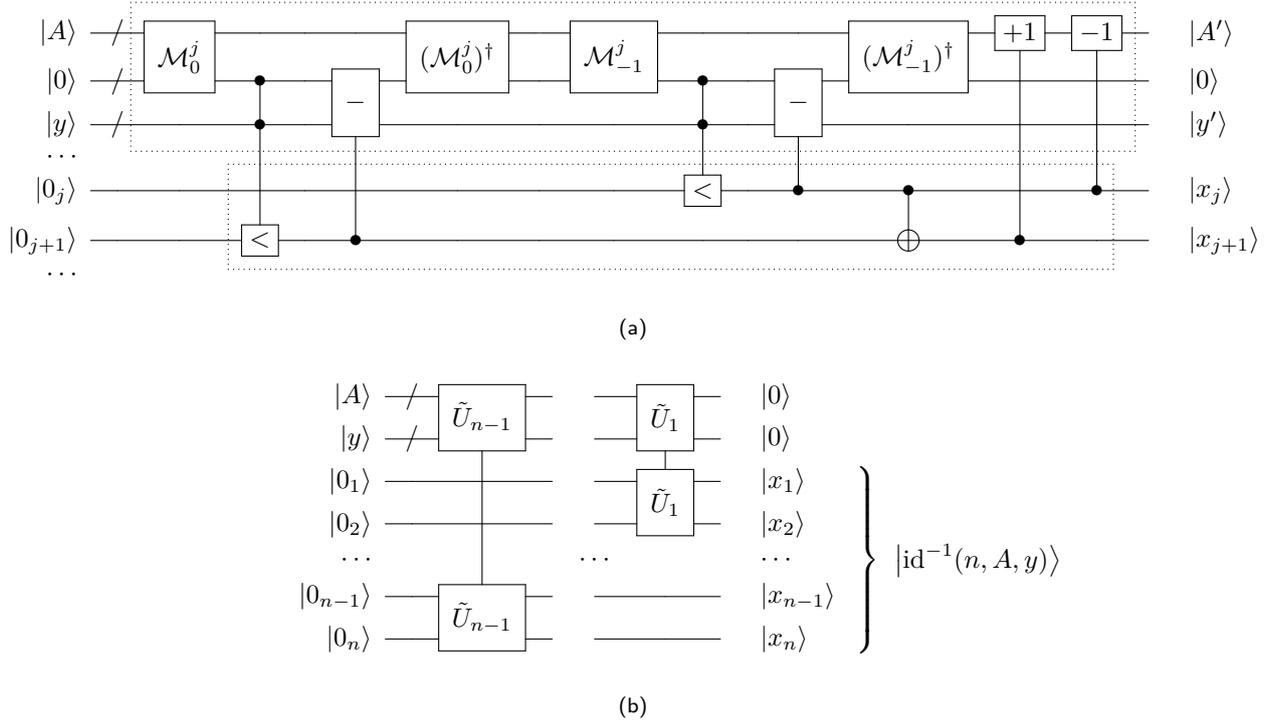

	The quantum circuit shown in \cref{fig:qcircuit} performs the unitary mapping $U_\#^\dag \ket{j} = \ket{\text{id}^{-1}(n, A, j)}$, i.e. un-indexes an integer to the corresponding portfolio. The correctness of the circuit follows directly from \cref{alg:unindex}, where we use the property that $y=0$ and $A=0$ at the end of the recursive sequence to ensure that there are no registers entangled with the output. Clearly by reversing the circuit, the indexing unitary $U_\#$ is obtained. 
	In the sub-circuit shown in \cref{fig:sub-first}, we rely on a unitary implementation of the counting function $\mathcal{M}^j_{k} \ket{A}\ket{0} = \ket{A} \ket{\mathcal{M}(j, A+k)}$. {Since $j$ is known classically for each sub-circuit, and we must have $-j \leq A+k \leq j$, this can be implemented by classically pre-computing the unique possible values of $\mathcal{M}(j, A+k)$ and applying at most $j+1=\mathcal{O}(n)$ controlled additions of these possible values (where the control is on equality with each possible value of $A+k$).
	In addition, the given indexing sub-circuit uses $O(1)$ quantum comparators and other controlled subtraction/addition operations. This leads to a sub-circuit complexity of $\mathcal{O}(n)$ as per {\cref{fig:sub-first}} and thus an overall indexing gate complexity of $\mathcal{O}(n^2)$ as per {\cref{fig:sub-second}}, omitting polylogarithmic factors.}
	
    The quantum circuit for quantum portfolio indexing relies on three main registers. An ancilla register of size $\mathcal{O}(\log n)$ is used to track the value of $A$ through the indexing process. The input index is held in a register of size $\mathcal{O}(\log \mathcal{M}(n, A)) = \mathcal{O}(n)$, and the output portfolio encoding is also $\mathcal{O}(n)$. Note that the second ancilla register shown in \cref{fig:sub-first} is not required and is shown for graphical convenience -- the un-set bits of the output register can be used instead.
    
    With the indexing unitary in hand, the $\mathcal{M}$ valid portfolios can be `connected' using any $\mathcal{M}$-vertex graph over which an efficient quantum walk can be implemented. In particular, \cite{marshqwoa} describes an efficient approach for using an arbitrary circulant graph as a mixer, since this class of graphs can be simulated efficiently using the Quantum Fourier Transform \cite{ qiang2016efficient, mahasinghe2016efficient, Zhou2017qft, Zhou2017, Loke2017a, Loke2017b, Qiang2018, visualtracking2019}. In this work, we choose the complete graph, $K_\mathcal{M}$, due to its efficiency of implementation and global symmetry. {Details of the circuit for quantum walk over $K_\mathcal{M}$ can be found in {\cite{marshqwoa,QFT,AhokasQFT}}, having asymptotic gate complexity $\mathcal{O}(n)$ to simulate the walk with exponential precision, again omitting polylogarithmic factors.} Thus, we have
    \begin{equation}
        U_\text{QWOA}(t) = U_\# e^{-i t K_\mathcal{M}} U_\#^\dag \, ,
    \end{equation}
    {where the depth of the mixing unitary is dominated by that of the indexing procedure, leading to an overall complexity of $\mathcal{O}(n^2)$.}
    The associated equal superposition initial state is
    \begin{equation}
        \ket{\psi_0} = \frac{1}{\sqrt{\mathcal{M}}}U_\#^\dag \sum\limits_{x=0}^{\mathcal{M}-1} \ket{x} \, .
    \end{equation}
    
 	\subsubsection{Comparison of mixing circuits}
	
\begin{figure*}
        \centering
        \begin{subfigure}{.32\textwidth}
          \centering
          \includegraphics[width=\textwidth]{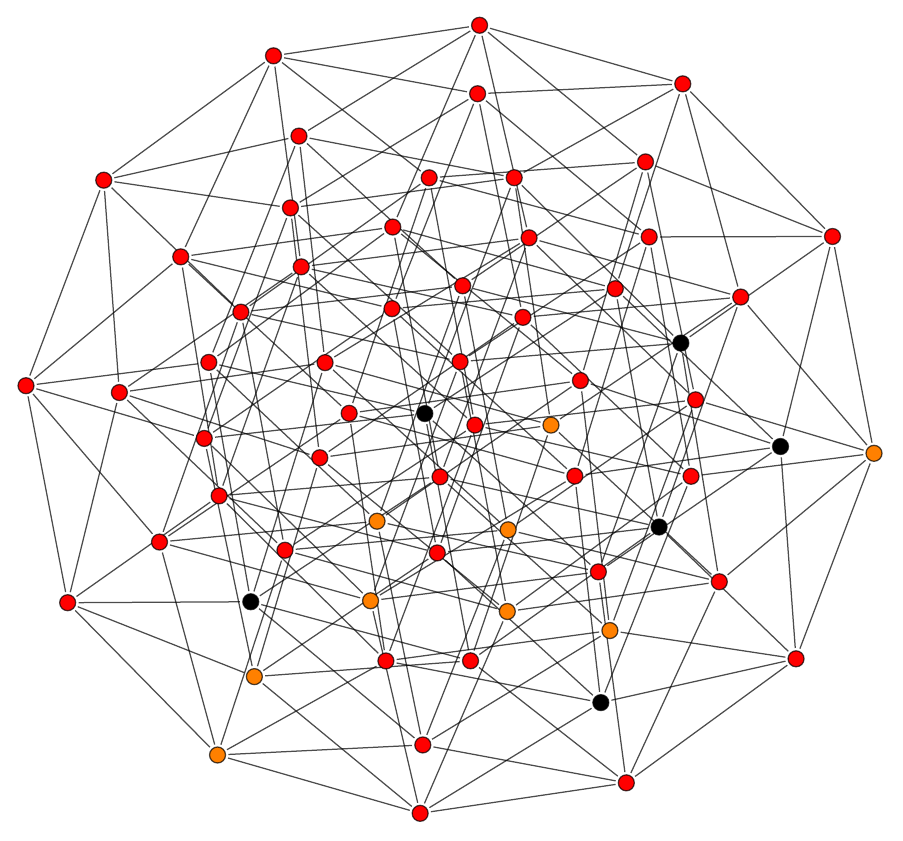}
          \caption{}
          \label{fig:adj-qaoa}
        \end{subfigure}
        \begin{subfigure}{.32\textwidth}
          \centering
          \includegraphics[width=\textwidth]{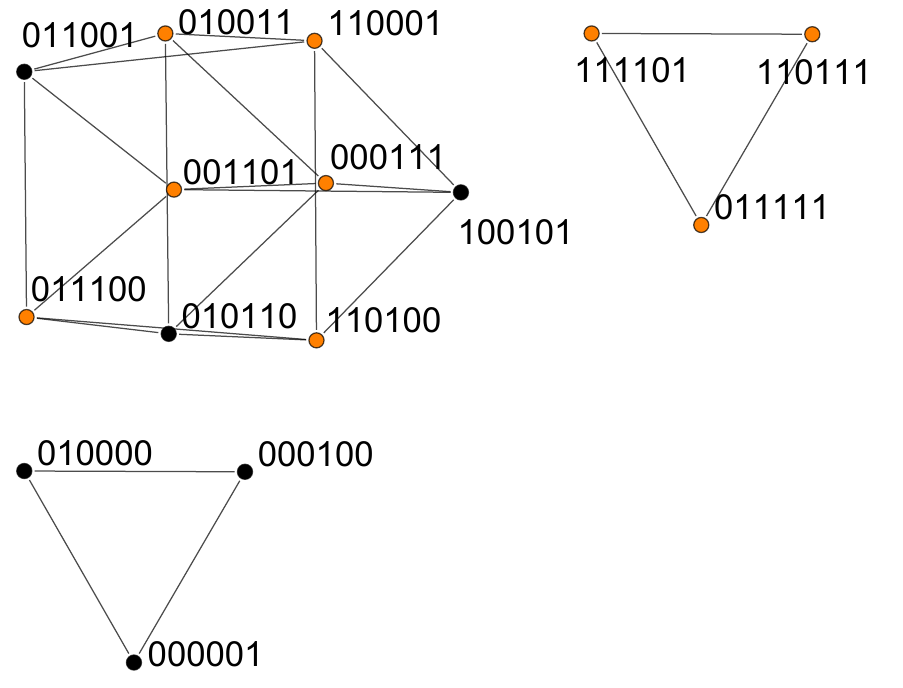}
          \caption{}
          \label{fig:adj-qaoaz}
        \end{subfigure}
        \begin{subfigure}{.32\textwidth}
          \centering
          \includegraphics[width=\textwidth]{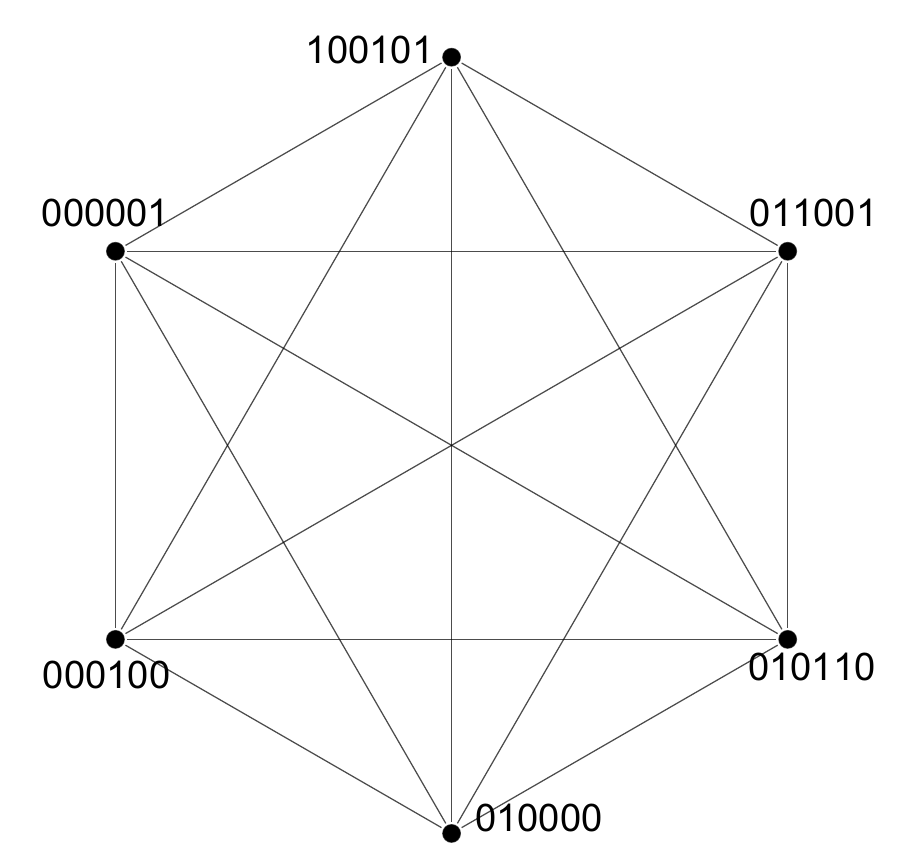}
          \caption{}
          \label{fig:adj-qwoa}
        \end{subfigure}
    \caption{An example $(n=3,A=1)$ of the graph representation of each underlying mixing Hamiltonian. Black vertices are valid non-degenerate solutions, orange vertices are degenerate solutions, and red vertices correspond to invalid portfolio configurations. (a) QAOA operates on the full $2^6$-dimensional Hilbert space, connected as a hypercube. (b) QAOAz uses the dual parity ring mixer on long and short qubits, and thus each graph component connects solutions of equal Hamming weight. (c) QWOA mixes uniformly between valid non-denegerate solutions, with full symmetry and transitivity amongst vertices and edges.}
    \label{fig:adjacency}
    \end{figure*}

	In \cref{fig:adjacency}, we represent the Hamiltonian underlying each mixing operator as the adjacency matrix of a graph for a 3-asset example problem, to illustrate the connectivity between solutions. The generic QAOA mixing operator can be considered as a quantum walk on the $2n$-dimensional hypercube shown in \cref{fig:adj-qaoa}. 
	QAOAz mixes only valid solutions, but as per \cref{fig:adj-qaoaz} there are a number of disconnected graph components, asymmetry in vertex degree and connectivity, and degeneracy amongst some of the valid portfolio solutions. Finally, associated with QWOA is a complete graph connecting the canonical valid solutions.
	
\begin{figure}[h!]
        \centering
        \begin{subfigure}{.45\textwidth}
          \centering
            \begin{tabular}{|l|l|l|l|l|}
            \hline
                            & QAOA             & QAOAz            & QWOA                \\ \hline
            Search space    & $2^{2n}$         & $\binom{2n}{n+A}$ & $\mathcal{M}(n, A)$ \\ \hline
            Gate complexity & $\mathcal{O}(n)$ & $\mathcal{O}(n)$  & $\mathcal{O}(n^2)$  \\ \hline
            \end{tabular}
            \caption{}
          \label{fig:compare-first2}
        \end{subfigure}
        \hfill
        \begin{subfigure}{.45\textwidth}
          \centering
          \includegraphics[width=\textwidth]{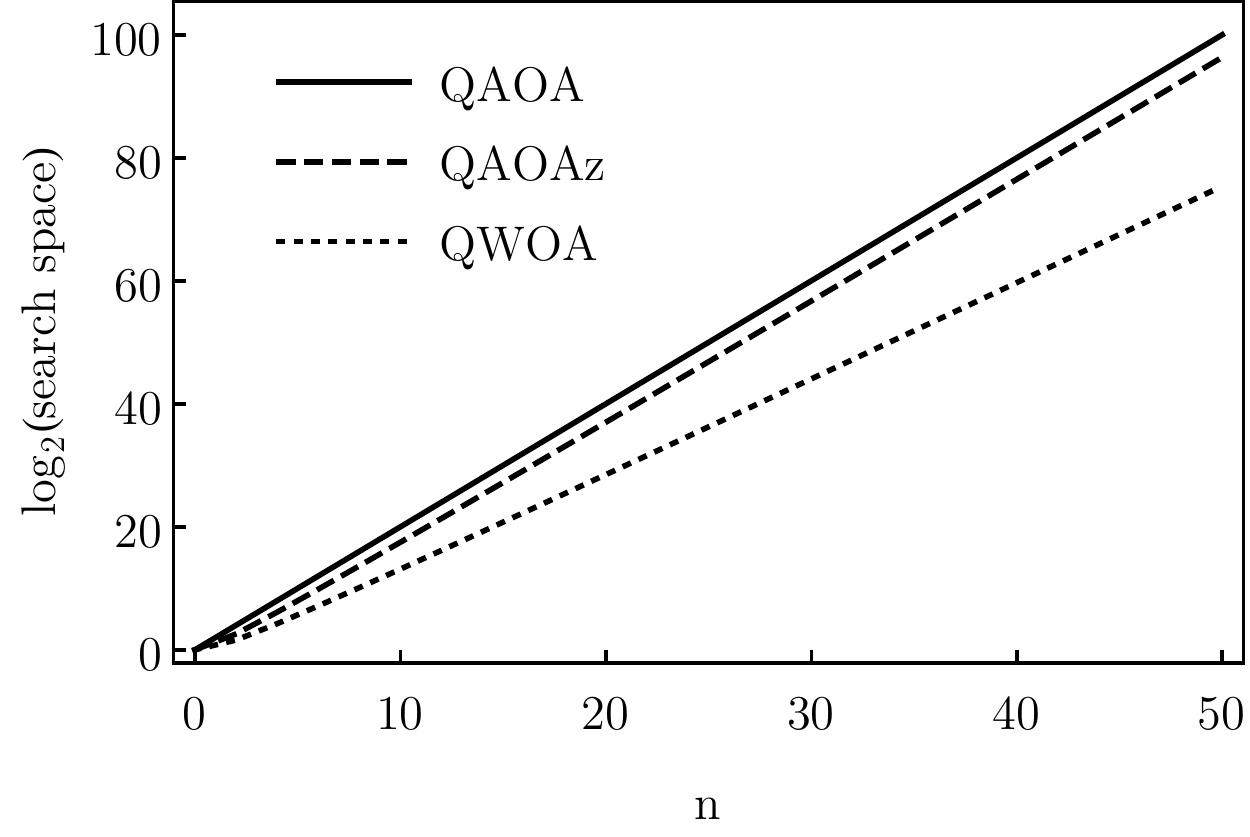}
          \caption{}
          \label{fig:sub-first2}
        \end{subfigure}
        \caption{(a) comparison of the mixing operators for the three approaches, {where the QWOA complexity is derived in {\cref{sec:finance-qwoa-method}}}. (b) comparison of the search space size for $A=0$.}
        \label{fig:comparemethods}
    \end{figure}

    In \cref{fig:comparemethods} we contrast the search space and gate complexity for the three approaches. Also shown is a plot of the search space for varying $n$, with fixed $A=0$. QAOA scales like $4^{n}$, while QAOAz scales like $\frac{4^{n}}{\sqrt{\pi n}}$ for large $n$. QWOA in contrast reduces the search space by more than half on mean, scaling approximately as $\frac{3^n}{2\sqrt{\pi n}}$ (this result is obtained by observing that for even $n$, $\mathcal{M}(n, 0)$ are the sum of squared trinomial coefficients). We argue that the significant reduction in the size of the search space is worth the quadratic increase in mixing circuit complexity as per \cref{fig:compare-first2}. As shown in the following section, QWOA converges to high-quality solutions far more quickly, with a small choice of $p$ producing near-optimal solutions.
    
    \section{Numerical Results and Analysis}
    
 Numerical simulation of the portfolio optimisation problem was carried out for QAOA, QAOAz and QWOA, using the software package QuOP\_MPI \cite{QuopEdric,matwiejew2020qswmpi}. For all reported results, the $\vec{\gamma}$ and $\vec{t}$ parameters were optimised using the Broyden-Fletcher-Goldfarb-Shanno algorithm. The Newton-Conjugate-Gradient, Nelder-Mead Simplex and Powell's method algorithms were also considered. However, these yielded equivalent, or uniformly poorer, results.

    The simulations utilised daily share prices from two data sets, Data Set A and B, which selected $n=8$ stocks from the ASX.20 index. The adjusted close price was used to ensure that all possible corporate actions were considered in the share performance, such as dividends. Data Set A contained the daily adjusted close prices from 01/01/2017 to 31/12/2018 for the shares: AMP, ANZ, AMC, BHP, BXB, CBA, CSL and IAG. Data Set B ranged over the period of 24/03/2020 to 06/09/2020 and selected shares based on sectors heavily impacted by the COVID-19 pandemic: FLT, QAN, WEB, REX, AIZ, SYD, SCG and CTD. Both data sets were obtained from Yahoo Finance using a Python based API \cite{pandas}. For all simulations, $A=4$ and $\lambda=0.5$. In each case, the presented results correspond to the mean and standard deviation for 15 repeats of each algorithm. These used the same set of randomly generated initial $\vec{\gamma}$ and $\vec{t}$ values, uniformly distributed between $0$ to $2\pi$. 
 
\begin{figure}[h!]
        \centering
        \begin{subfigure}{.45\textwidth}
          \centering
          \includegraphics[width=\textwidth]{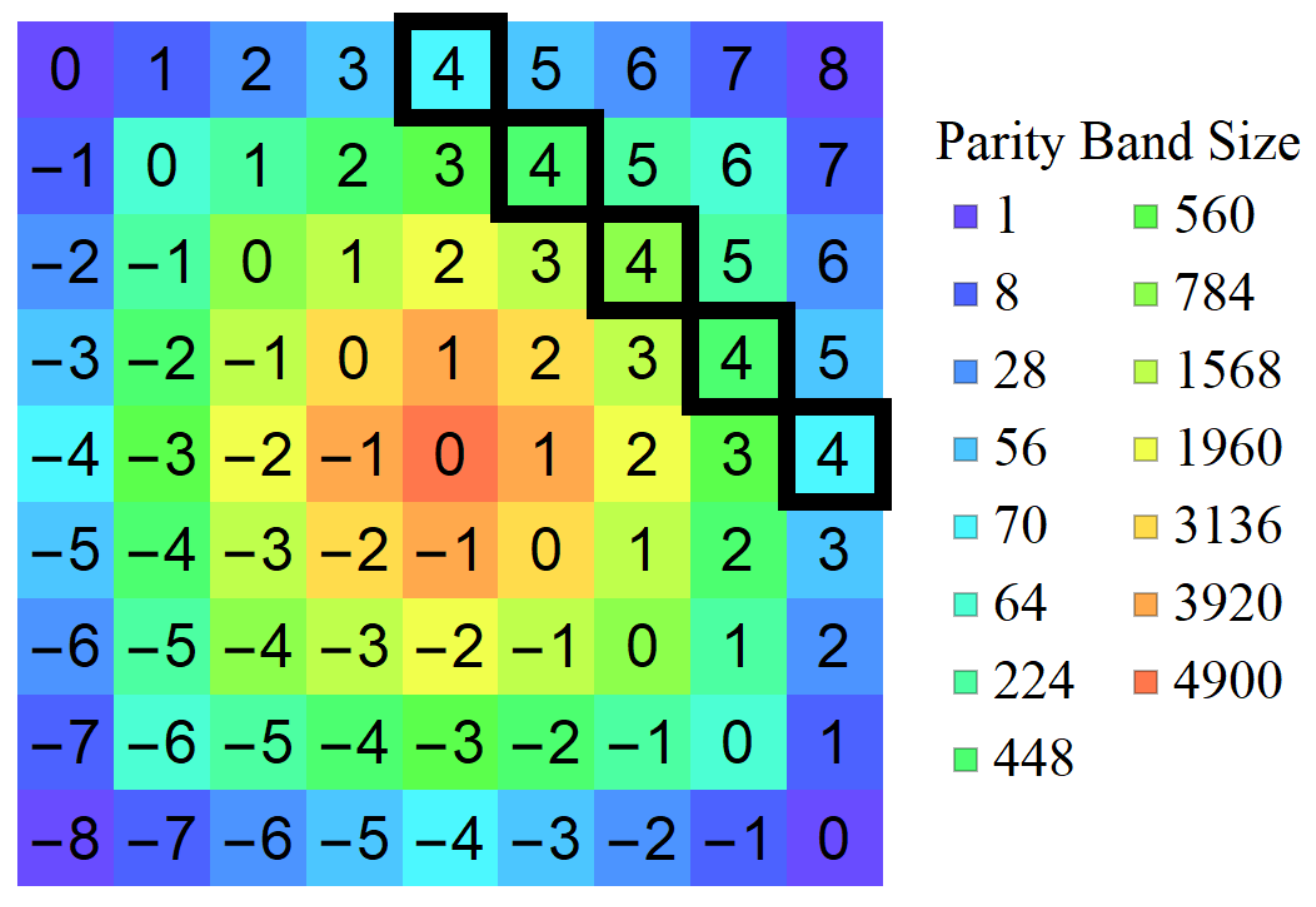}
          \caption{}
          \label{fig:parity_subgraphs}
        \end{subfigure}
        \hfill
        \begin{subfigure}{.45\textwidth}
          \centering
`          \includegraphics[width=\textwidth]{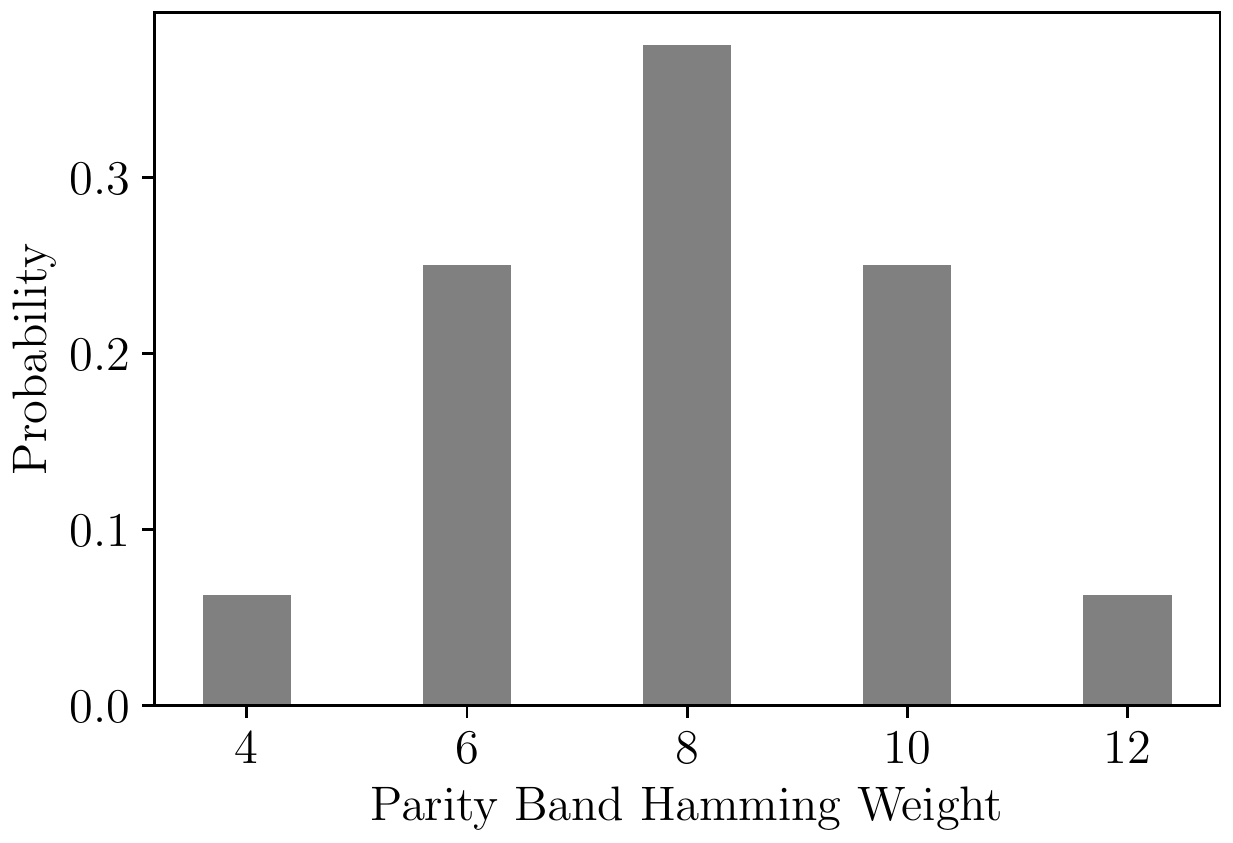}
          \caption{}
          \label{fig:parity_state_distribution}
        \end{subfigure}
        \caption{(a) QAOAz parity band size and associated constraint value, $A$, for $N=8$ assets. Parity bands satisfying $A = 4$ are outlined in black. (b) Probability distribution of $\ket{\psi_0}$ over the QAOAz parity bands for $A=4$ as a function of Hamming weight. This distribution occurs across the $5$ outlined parity bands in \cref{fig:parity_subgraphs} where the Hamming weight increases down the outlined diagonal.}
        \label{fig:parity_mixer_distribution}
    \end{figure}

    With $N=8$ and $A=4$, the search spaces of the three algorithms are $2^{16}$ for QAOA, 1820 for QAOAz and 266 for QWOA. The size of the disconnected QAOAz parity bands are shown in \cref{fig:parity_subgraphs} for $N=8$ along with the distribution of $\ket{\psi_0}$ across all bands satisfying $A=4$. We note that the largest connected components correspond to states satisfying $A$ equal or close to $0$, and that the binomial distribution of the initial state centers on these larger components. As mixing of probability amplitude does not occur across the parity bands, the probabilities shown in \cref{fig:parity_state_distribution} are the maximum possible for a single state in each of the parity bands, as opposed to QAOA and QWOA which converge to a single state as $p \rightarrow \infty$. For Data Set A, this influences the minimum possible objective function value which is $-0.318$ for QAOA and QWOA, and $-0.305$ for QAOAz. For Data Set B the minimum is $-1.25$ for QAOA, QWOA and QAOAZ as the optimal portfolio exists in all four of the populated QAOAz parity bands. The small and fully-connected search space of QWOA is expected to result in it outperforming QAOA and QAOAz at sufficiently high $p$.
    
    Of critical note is that the classical optimisation component suffers from the curse of dimensionality as the number of optimisation parameters increase (i.e. with increasing $p$). Each time $p$ increases by 1, two additional independent classical optimisation parameters are needed to maximally explore the available search subspace. In hybrid quantum-classical optimisation an estimation of $\bra{\psi_f} C \ket{\psi_f}$ is obtained by repeated sampling of the $\ket{\psi_f}$ state, and the optimisation parameters are tuned to optimise this quantity. Consequently, with increasing dimensionality the classical optimisation deteriorates \cite{CurseOfDimensionality}, making less progress per optimisation iteration, and thus increasing the overall number of quantum circuit shots required. A practical hybrid quantum-classical variational algorithm must therefore demonstrate rapid convergence at low $p$. In this sense, the polynomial difference in circuit depth per iteration given in \cref{fig:comparemethods} is negligible compared to the exponentially increasing classical parameter space with increasing $p$. In the following numerical results we demonstrate that QWOA is by far the best candidate in this regard out of the three considered algorithms, needing drastically smaller $p$ to reach a given expected solution quality $\bra{\psi_f} C \ket{\psi_f}$. 
    
    \subsection{Data Set A (01/01/2017 to 31/12/2018)}\label{seta}
  
\begin{figure}[h!]
        \centering
        \begin{subfigure}[b]{.49\textwidth}
            \includegraphics[width=\textwidth]{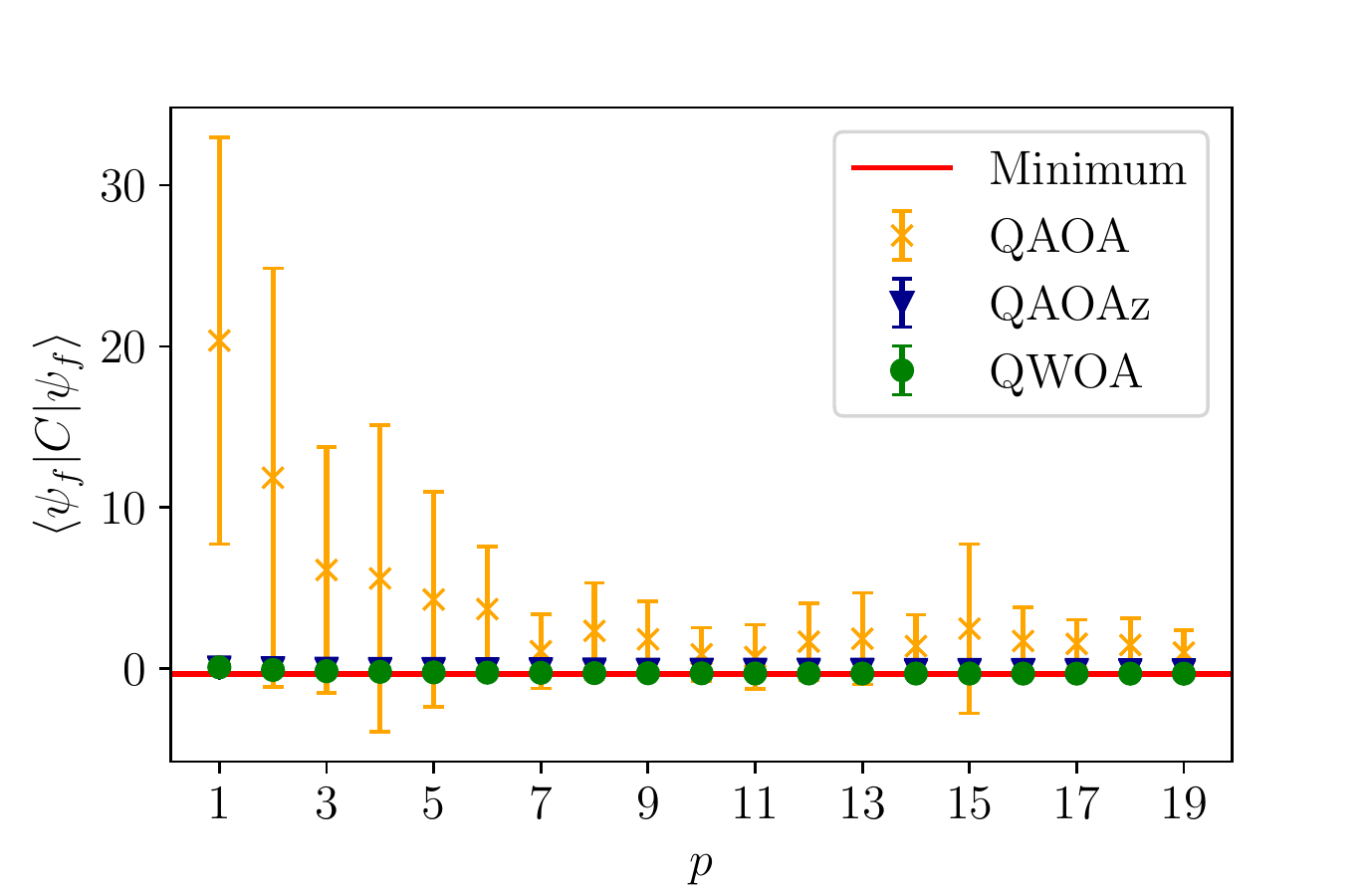}
            \caption{}
            \label{fig:allobjstand}
        \end{subfigure}
        \begin{subfigure}[b]{.49\textwidth}
            \includegraphics[width=\textwidth]{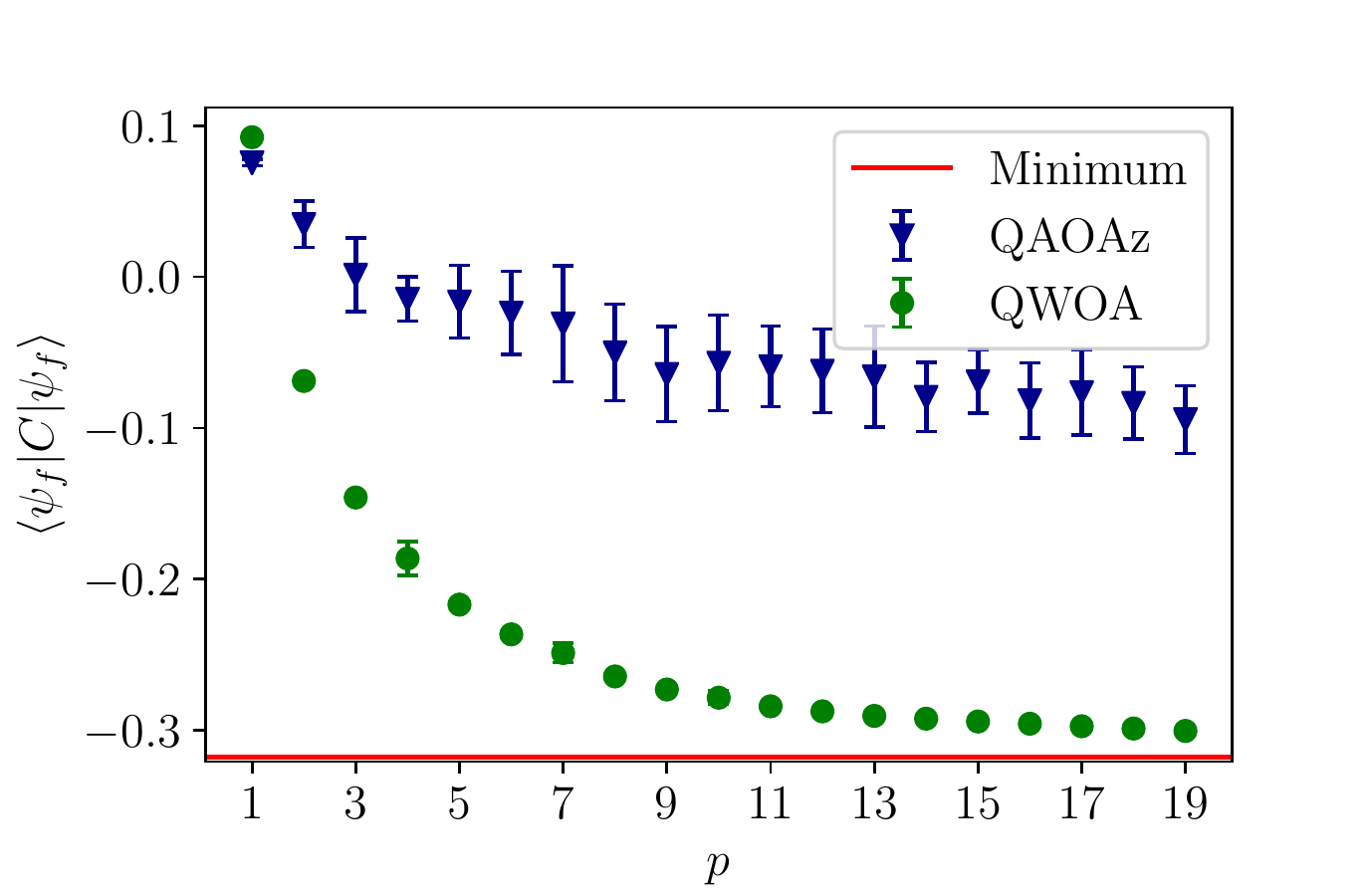}
            \caption{}
            \label{fig:standzoom}
        \end{subfigure}
        \caption{Expected portfolio quality for Data Set A as a function of $p$. (a) Comparison of QAOA, QAOAz and QWOA. (b) Excluding QAOA.}
    \end{figure}

    \cref{fig:allobjstand} shows the final value of the optimised objective function for QAOA, QAOAz and QWOA. On average, QAOA performs poorly as compared to the other two algorithms. Additionally, QAOA consistently exhibits the largest standard deviation in the optimised objective function value $\bra{\psi_f}C\ket{\psi_f}$ with a maximum of $12.96$ as compared to $0.038$ for QAOAz and $0.011$ for QWOA. This is consistent with the inclusion of invalid portfolio states and the QAOA mixing operator's action over the complete Hilbert space, as per \cref{fig:comparemethods}, which increases the likelihood of the classical optimiser converging to a poor local minima when compared to QAOAz or QWOA. {Note that the minimum objective function value (red line), derived from the globally optimal portfolio selection, is obtained by classical brute-force iteration through all feasible solutions and shown for the purpose of comparison.}
    
    \begin{figure}[b!]
        \centering
        \begin{subfigure}{.49\textwidth} 
            \centering
            \includegraphics[width=\textwidth]{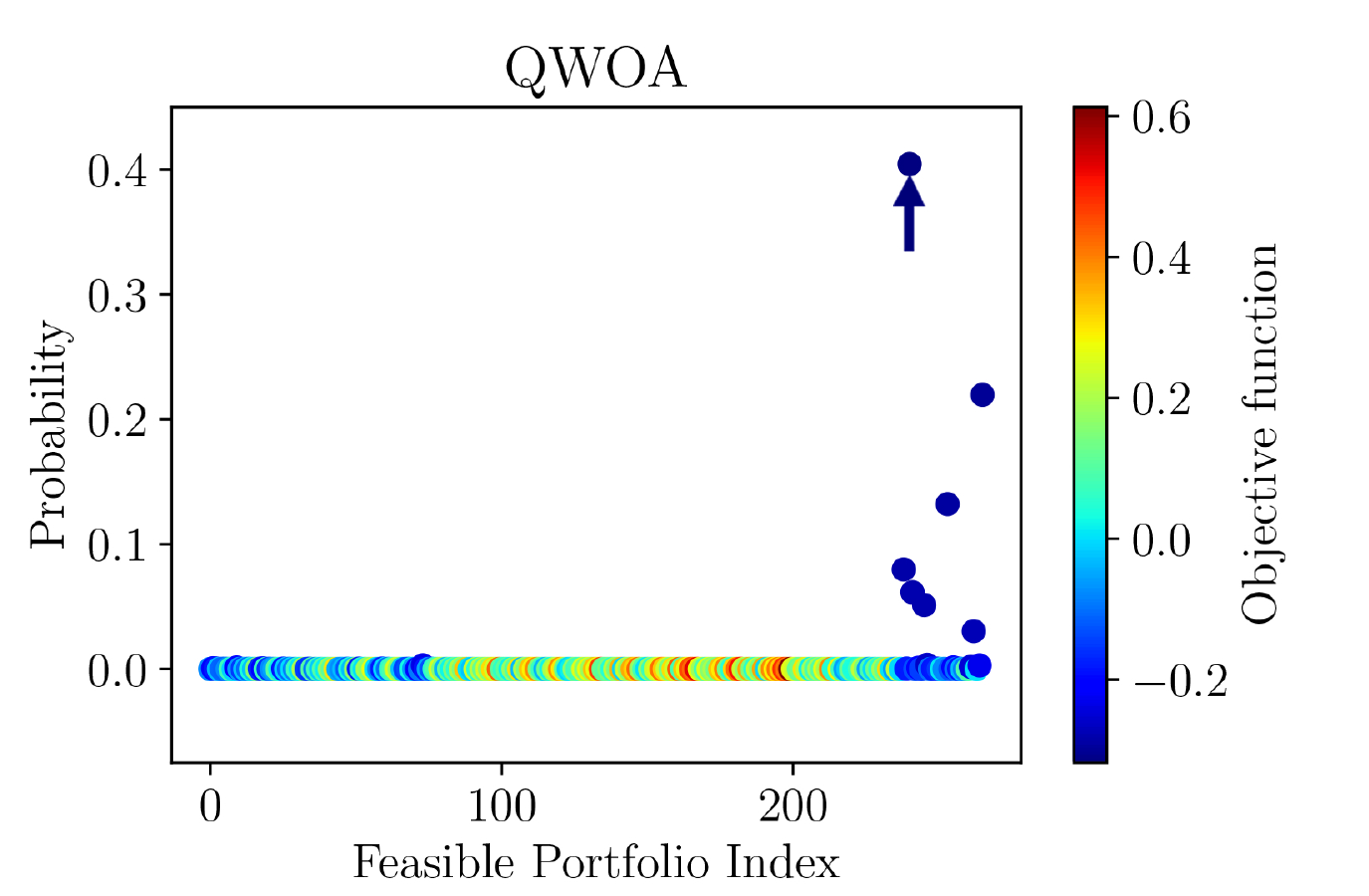}
            \caption{}
            \label{fig:statesa_qwoa}
        \end{subfigure}
        \begin{subfigure}{.49\textwidth} 
            \centering
            \includegraphics[width=\textwidth]{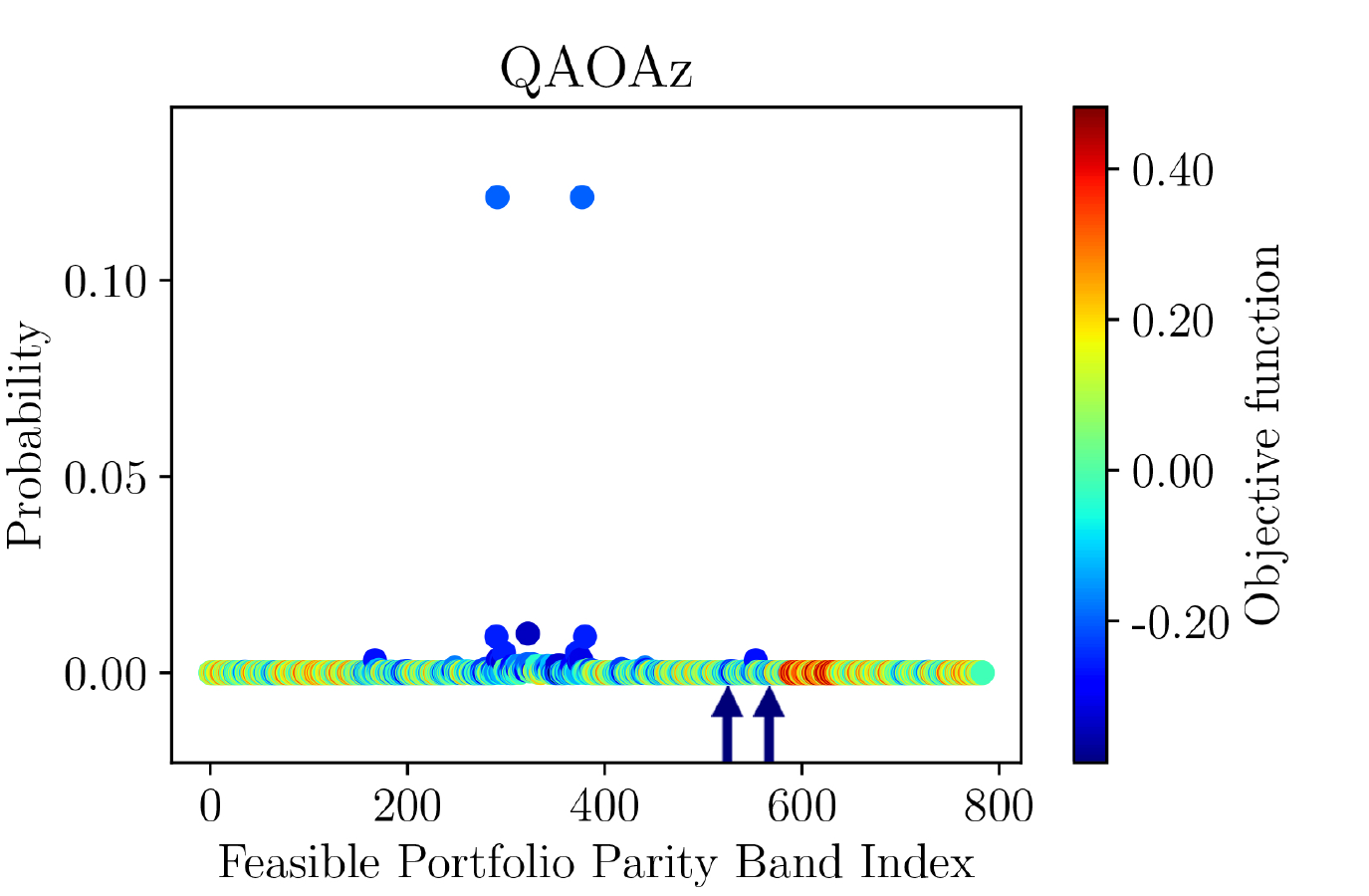}
            \caption{}
            \label{fig:statesa_qaoaz}
        \end{subfigure}
        \caption{Comparison of output state probabilities for QWOA and QAOAz using $p=19$ with Data Set A, with optimal solutions indicated by arrow. (a) QWOA amplifies the optimal portfolio to over 40\% probability, as well as boosting other high-quality portfolios. (b) QAOAz does not amplify the (degenerate) optimal solutions in this parity band, instead converging to solutions of comparatively lower quality.}
        \label{fig:statesa}
    \end{figure}
    
    The optimised expectation value after $p$ iterations for QAOAz and QWOA is shown in \cref{fig:standzoom}. QAOAz shows diminishing improvement past $p\approx8$, reaching an expected portfolio value approximately $0.2$ above the optimum portfolio objective function value after $19$ iterations. To explore the reason behind this performance, \cref{fig:statesa_qaoaz} plots the QAOAz output state probabilities for solutions contained in the parity band of Hamming weight 8, see also \cref{fig:parity_mixer_distribution}, which corresponds to a parity band containing the optimal solution. It is clear that degeneracy plays a part, with the two highest-probability states in the parity band representing the same portfolio. Even though this is a parity band containing the (degenerate) optimal solution state, its associated probability has not been amplified, remaining below $10^{-9}$. 
    Conversely, QWOA exhibits rapid improvement with $p$ over the range considered as per \cref{fig:standzoom} and, as shown in \cref{fig:statesa_qwoa}, is able to amplify the optimal portfolio configuration to above 40\% probability by $p = 19$.
        
    Overall, the superior performance of QWOA at low $p$ is consistent with the algorithm's reduction of the search space to 0.406\% of QAOA and 14.61\% of QAOAz; along with a reduction in state and mixing bias. Given $n=8$ with an investment constraint of $A=4$, there are $1554$ degenerate states, which are not equally distributed over the valid solutions. As per \cref{eq:QAOAz_size}, the number of the degenerate states for a given portfolio is directly related to the number of `no positions' it contains, with a higher number corresponding to more degenerate states. This clearly effects the performance of QAOA and QAOAz, producing lower-quality portfolios on average for a given choice of $p$.
    
\begin{figure}[h!]
        \centering
        \begin{subfigure}{.49\textwidth}
            \centering
            \includegraphics[width=\textwidth]{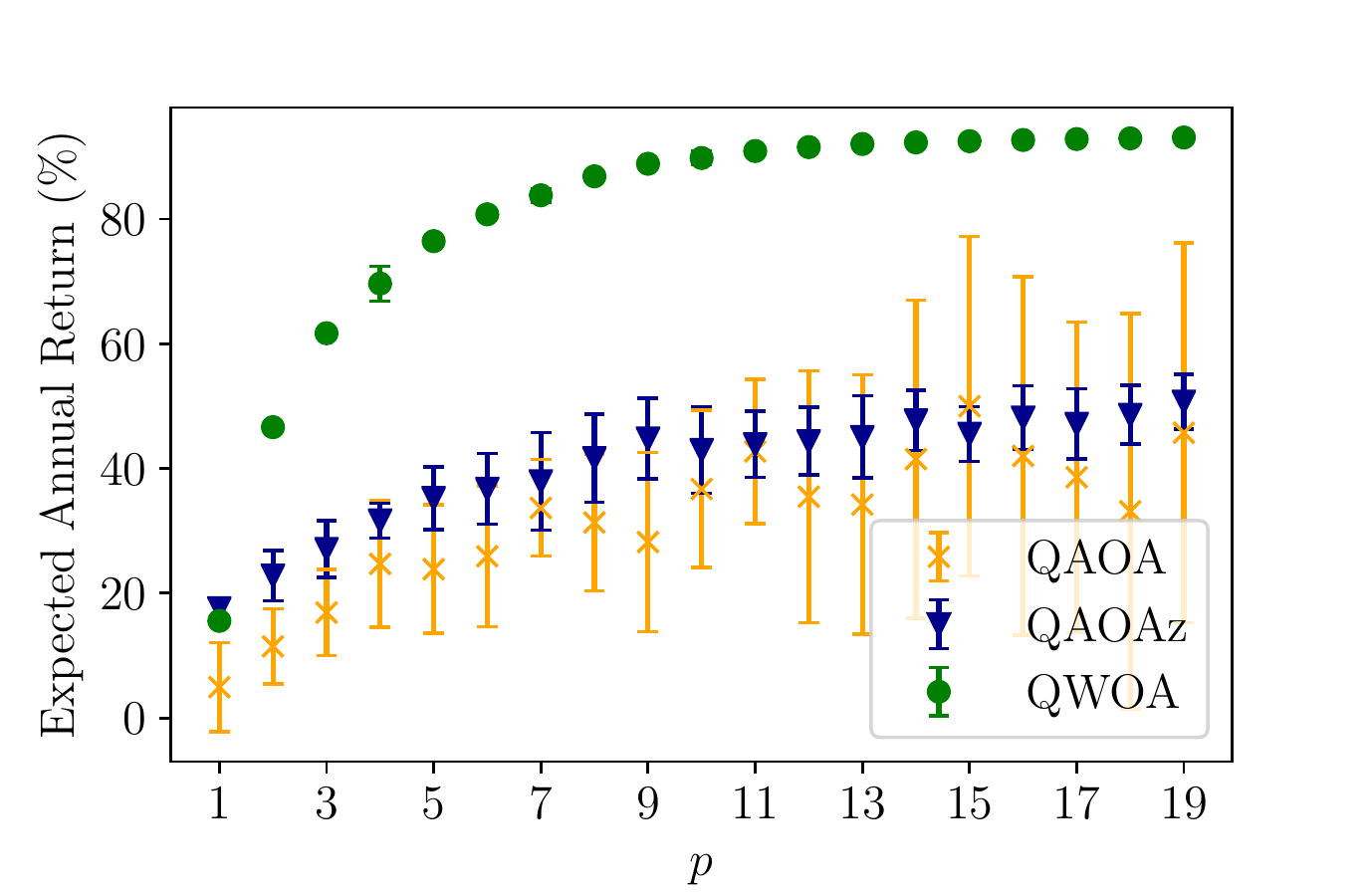}
            \caption{}
            \label{standexpecret}
        \end{subfigure}
        \begin{subfigure}{.49\textwidth}
            \centering
            \includegraphics[width=\textwidth]{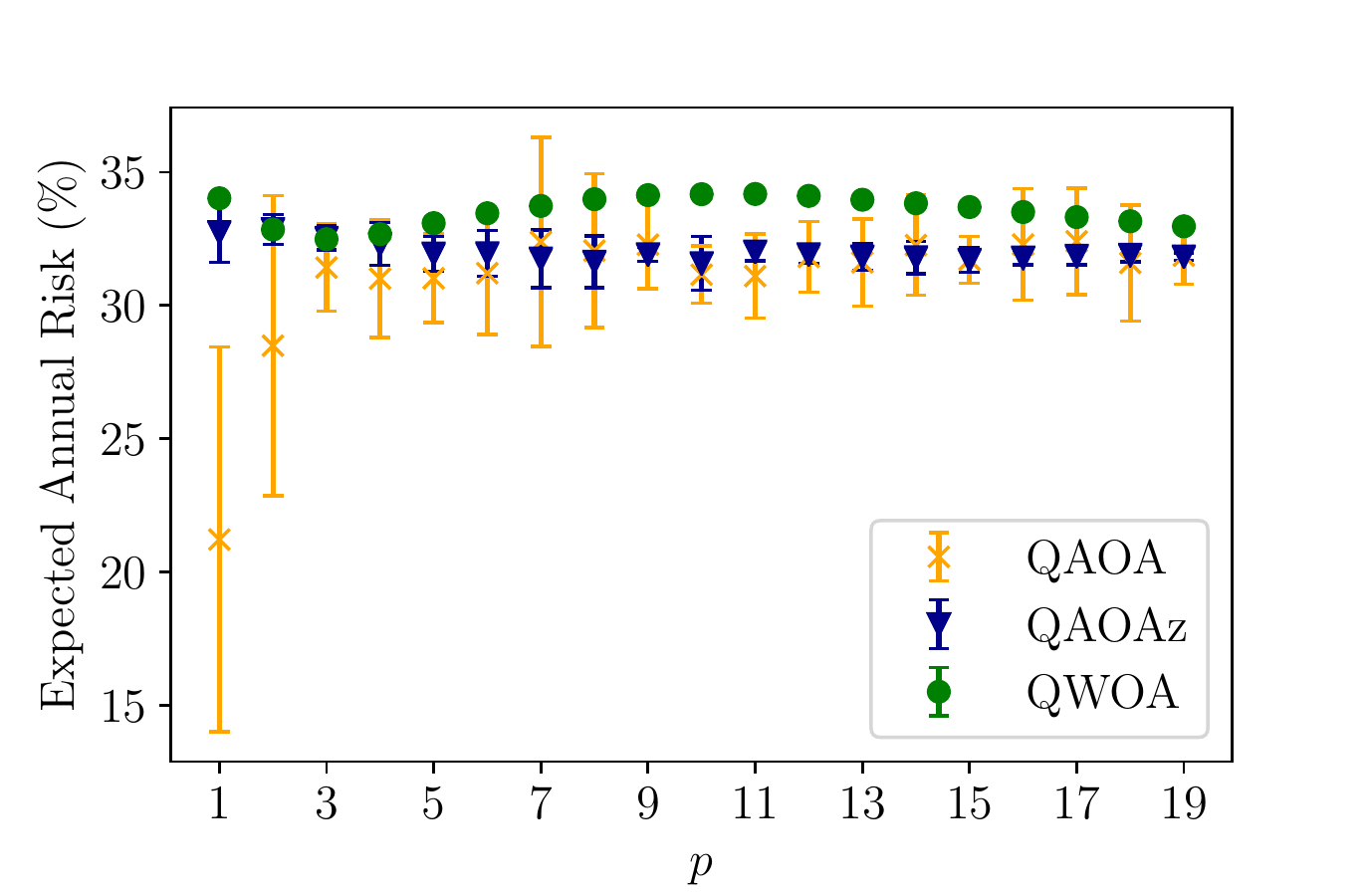}
            \caption{}
            \label{standexpecrisk}
        \end{subfigure}
        \begin{subfigure}{0.49\textwidth}
        \centering
            \includegraphics[width=\textwidth]{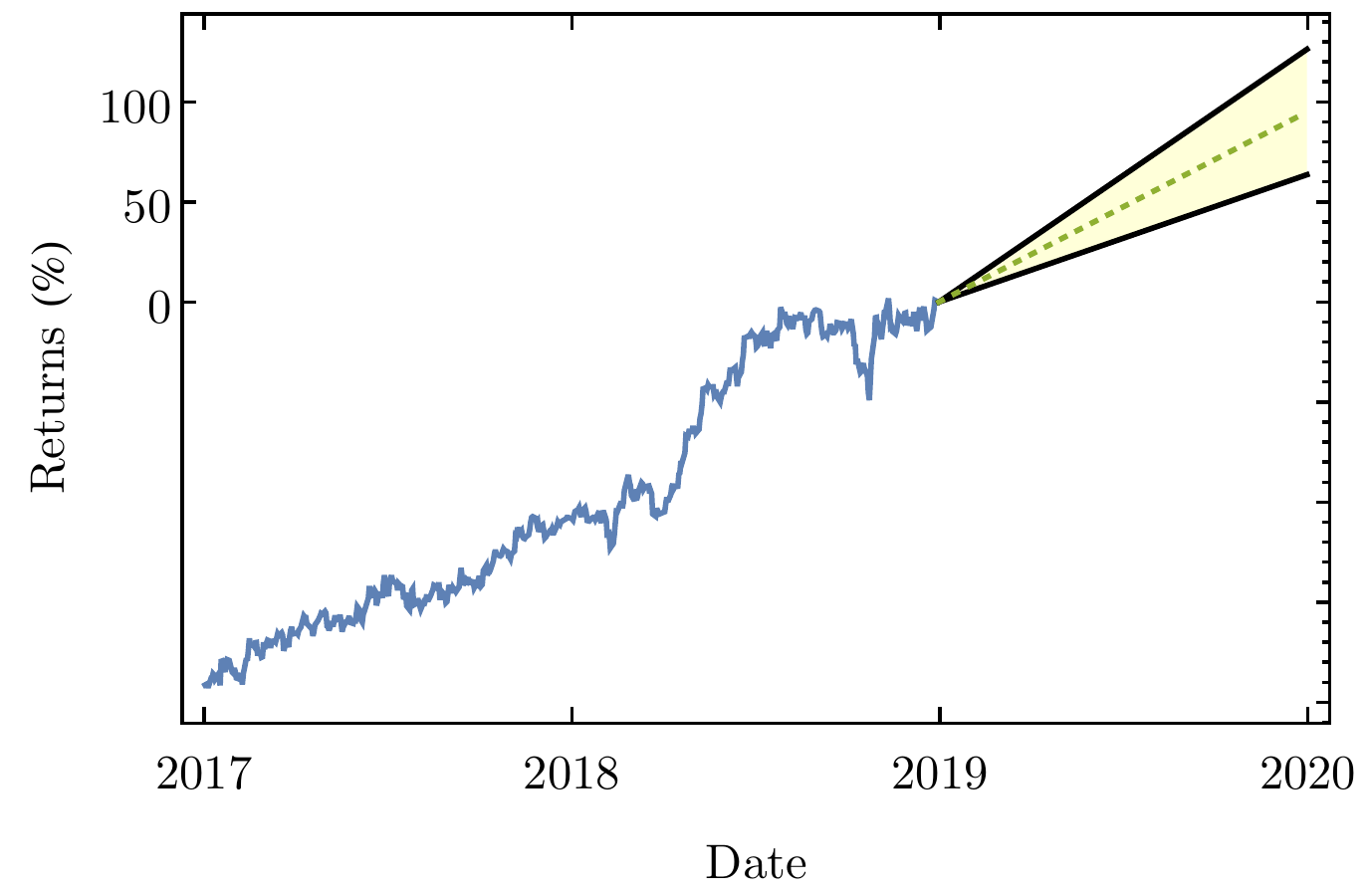}
            \caption{}
            \label{plot}
        \end{subfigure}
        \caption{Expected annual portfolio return (a) and expected annual risk (b) for the algorithms with increasing $p$ for Data Set A. The historic mean return for Data Set A is shown in (c) along with the projected expected return given by the final QWOA state at $p = 19$; with the yellow region being the 1-$\sigma$ risk region and the dotted line the expected value.}
        \label{fig:expectedret}
    \end{figure}

    \cref{standexpecret} and \cref{standexpecrisk} show the expected annual return and expected annual risk for Data Set A. These values were obtained by multiplying the probability of a each portfolio configuration by the corresponding annual return and risk respectively. QWOA performs significantly better than QAOA and QAOAz with respect to the annual return. We note that, as the mean-variance model is trying to find the best combination of return vs risk, the optimal portfolio will not necessarily exhibit the lowest possible risk - as low risk assets are typically associated with decreased return. \cref{plot} displays the historic mean return and the projected return of the final QWOA state at $p=19$.  It is clear that the obtained data is a good match for the historic data used, with an accurate representation of the trend and volatility of the data.
     
    \subsection{Data Set B (24/03/2020 to 06/09/2020)}\label{datab}
    
\begin{figure}[!h]
        \centering
        \begin{subfigure}[b]{.49\textwidth}
            \includegraphics[width=\textwidth]{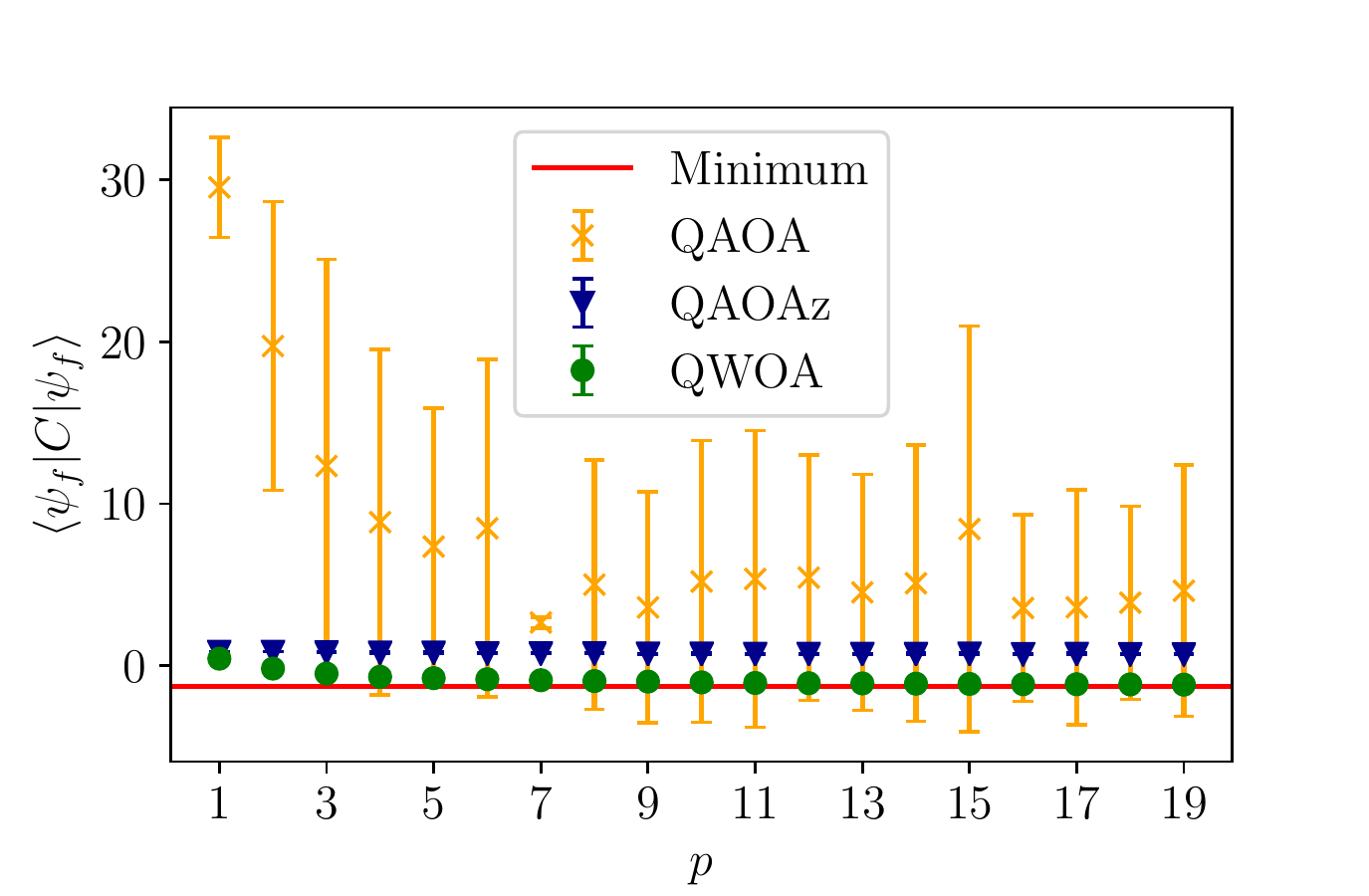}
            \caption{}
            \label{fig:allobj}
        \end{subfigure}
        \begin{subfigure}[b]{.49\textwidth}
            \includegraphics[width=\textwidth]{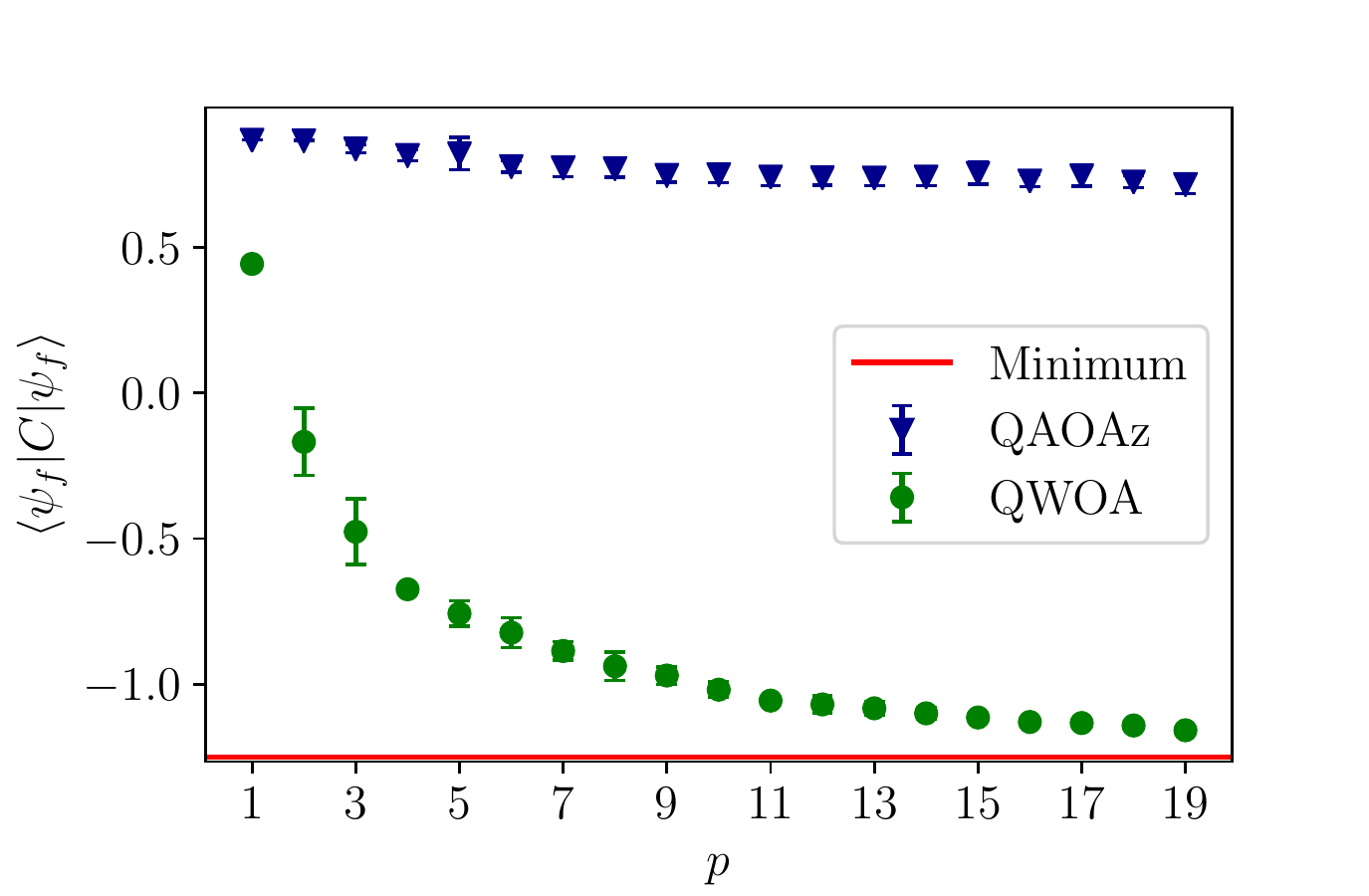}
            \caption{}
            \label{fig:hardQWsetb}
        \end{subfigure}
       \caption{Expected portfolio quality for Data Set B as a function of $p$. (a) Comparison of QAOA, QAOAz and QWOA. (b) Excluding QAOA.}
    \end{figure}

    Data Set B is consistent with the pattern of performance observed for Data Set A. As shown in \cref{fig:allobj} and \cref{fig:hardQWsetb}, QWOA consistently finds the best expected solution quality, followed by QAOAz and then QAOA. The same trend in the standard deviation of $\bra{\psi_f}C\ket{\psi_f}$ is also observed with QAOA having a maximum of $12.75$, followed by $0.61$ for QAOAz and $0.115$ for QWOA.

    As shown in \cref{fig:hardQWsetb}, QWOA again offers significant advantage over QAOAz. As previously discussed, the QAOAz parity mixer is expected to lead to performance variation dependant on the distribution of the optimal solutions amongst the disconnected parity bands. However, for Data Set B, the optimal solution is present in each of the 5 parity bands, as opposed to Data Set A, which contained the optimal solution in only $3$ of the 5 graphs (of sizes $448$, $784$ and $448$ respectively). A convergence to the optimal solution is observed in the QWOA numerical simulations, shown in \cref{fig:standzoom} and \cref{fig:hardQWsetb}.

\begin{figure}[h!]
        \centering
        \begin{subfigure}[t]{.49\textwidth}
            \includegraphics[width=0.99\textwidth]{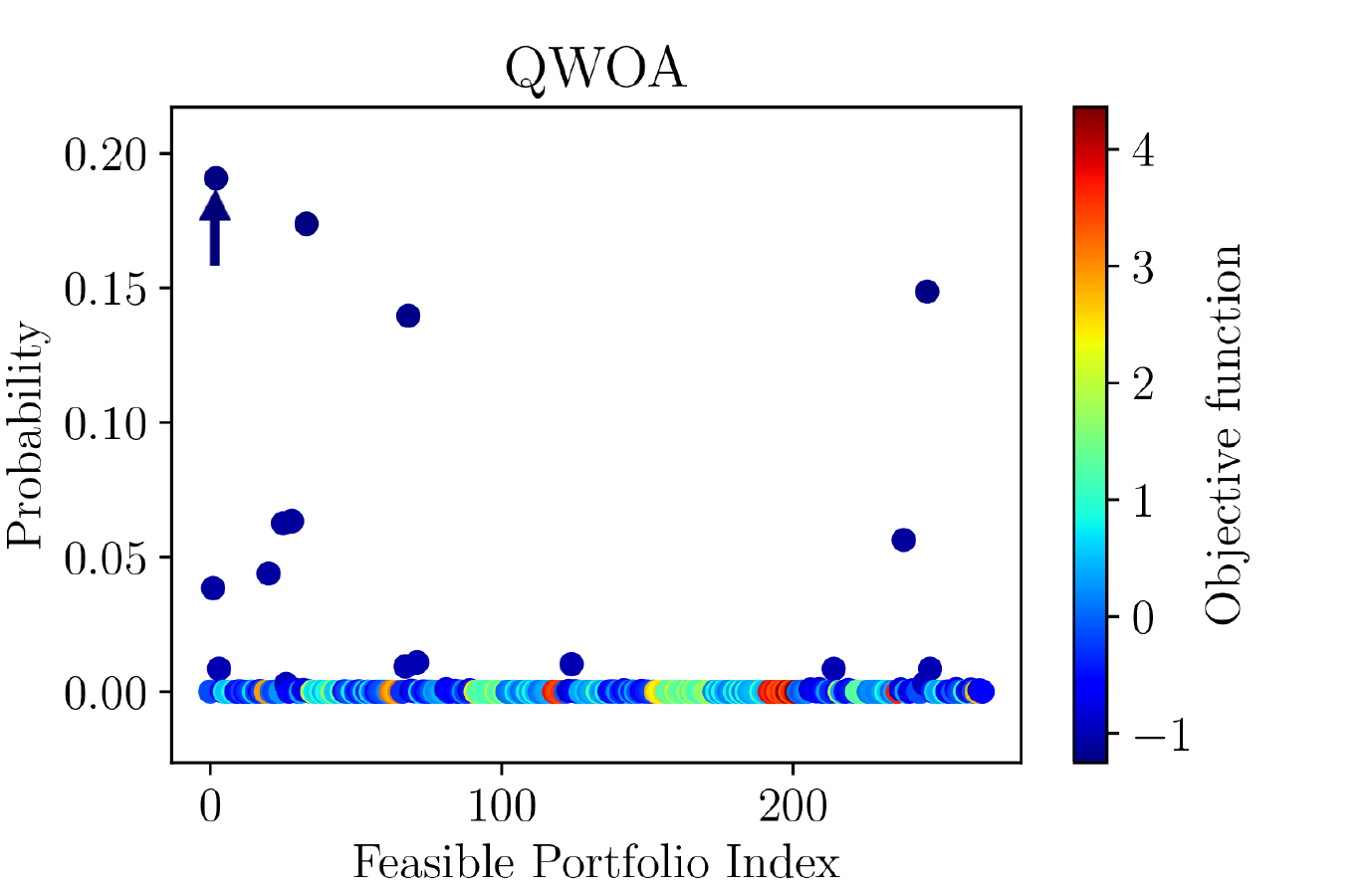}
            \caption{}
            \label{fig:statesb_qwoa}
        \end{subfigure}
        \begin{subfigure}[t]{.49\textwidth}
            \includegraphics[width=0.99\textwidth]{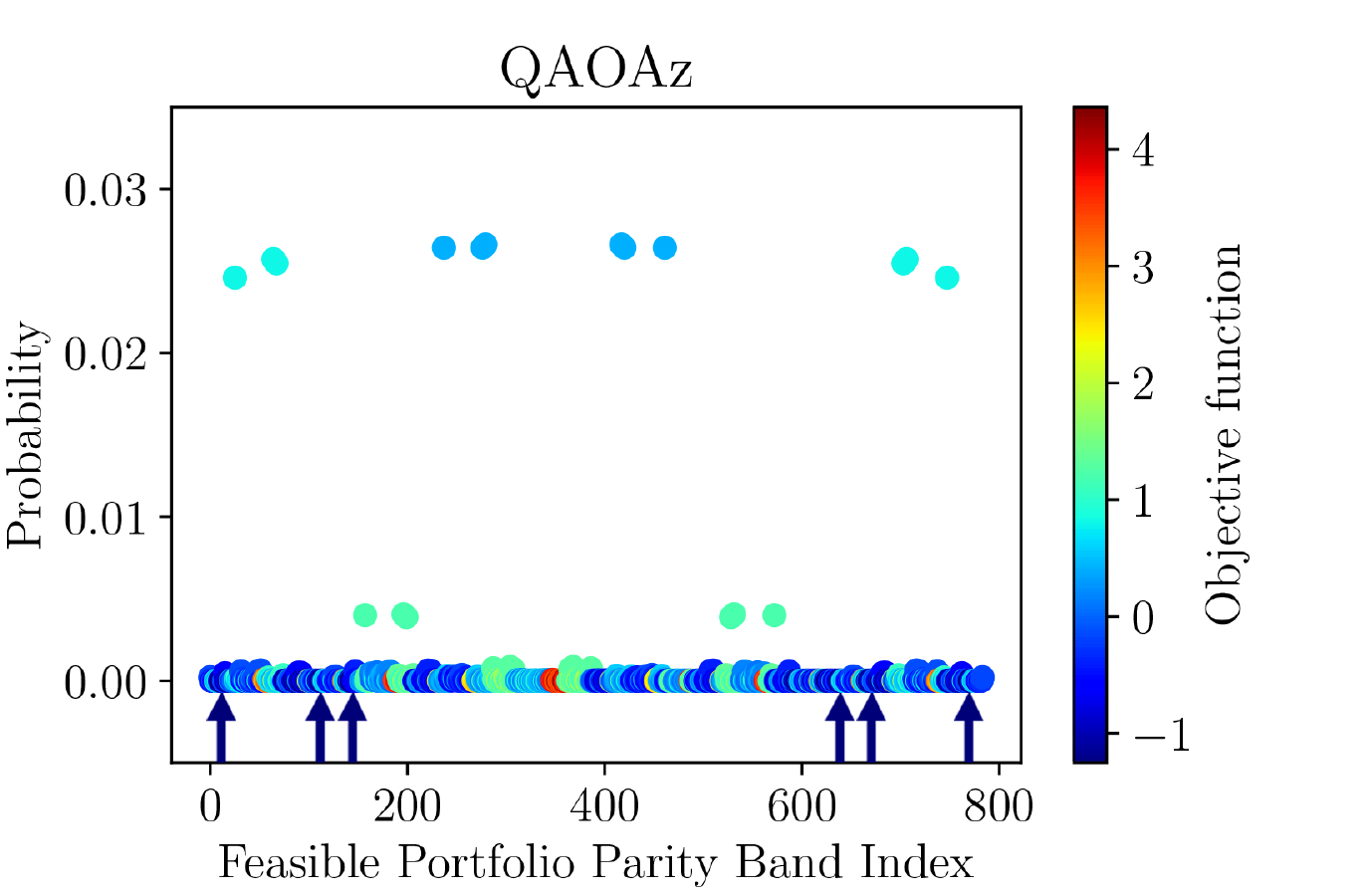}
            \caption{}
            \label{fig:statesb_qaoaz}
        \end{subfigure}
        \caption{Comparison of output state probabilities for QWOA and QAOAz using $p=19$ with Data Set B, with optimal solutions indicated by arrow. (a) QWOA amplifies the optimal portfolio to almost 20\% probability, as well as other high-quality portfolios. (b) QAOAz does not amplify the (degenerate) optimal solutions in this parity band, instead converging to solutions of comparatively lower quality.}
        \label{fig:statesb}
    \end{figure}

    Reinforcing the above observations, we note significant differences between QAOAz and QWOA in the probability distribution of the optimised state $\ket{\psi_f}$ for $p=19$, as shown in \cref{fig:statesb}. QWOA manages to boost the probability of the optimal portfolio state to approximately $20\%$. It also amplifies other high-quality portfolios, as demonstrated in \cref{fig:statesb_qwoa}. Examining the parity band shown in \cref{fig:statesb_qaoaz}, the optimised parameters amplify solutions of comparatively lower quality than the optimum with the probability for the degenerate optimal solutions remaining below $10^{-6}$.
    Also notable is the presence of solution degeneracy, with the 12 significantly amplified portfolios corresponding to four distinct degenerate solutions.
    
    \begin{figure}[h!]
        \centering
        \begin{subfigure}{.49\textwidth}
            \centering
            \includegraphics[width=\textwidth]{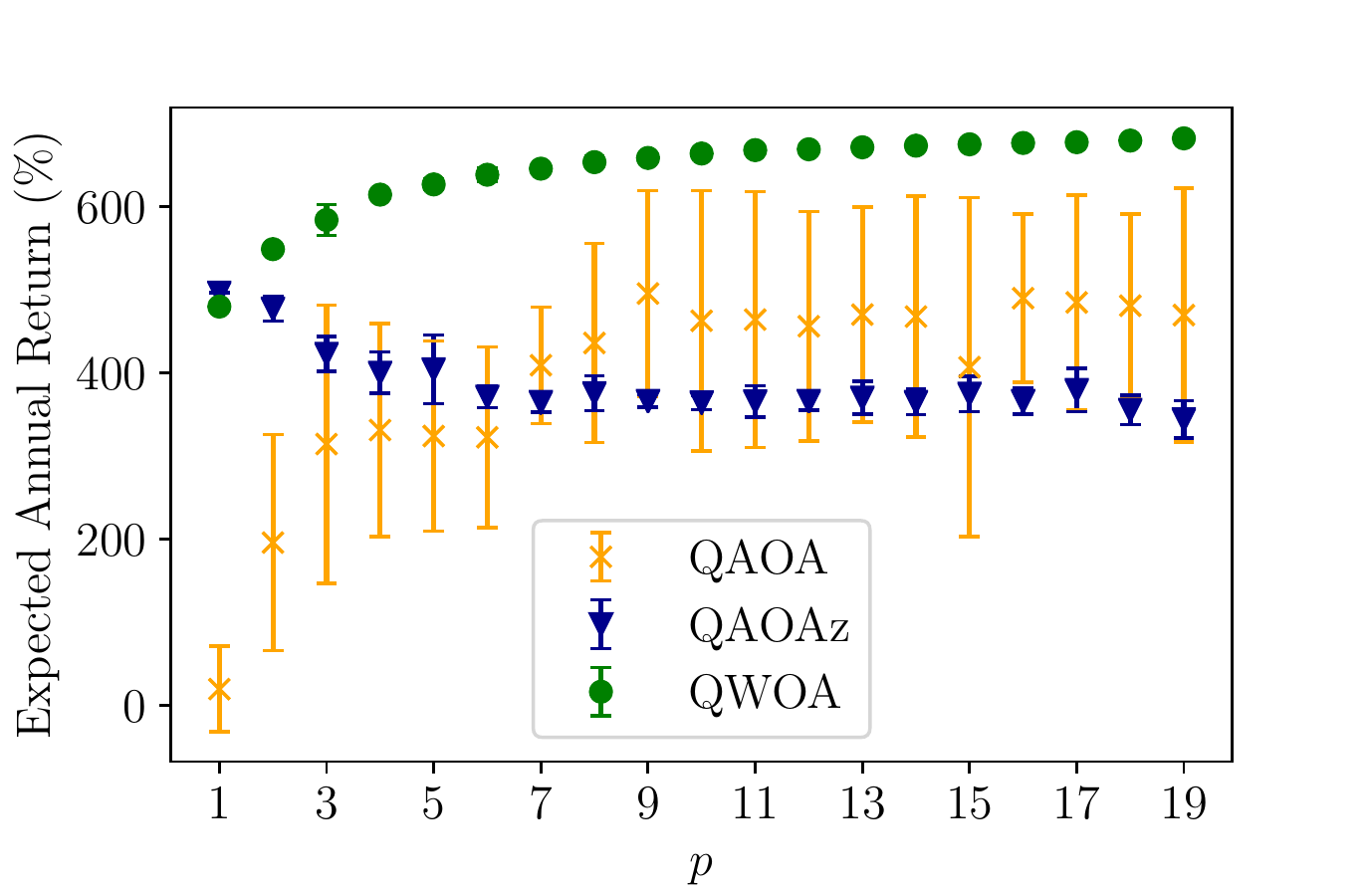}
            \caption{}
            \label{covidret}
        \end{subfigure}
        \begin{subfigure}{.49\textwidth}
            \centering
            \includegraphics[width=\textwidth]{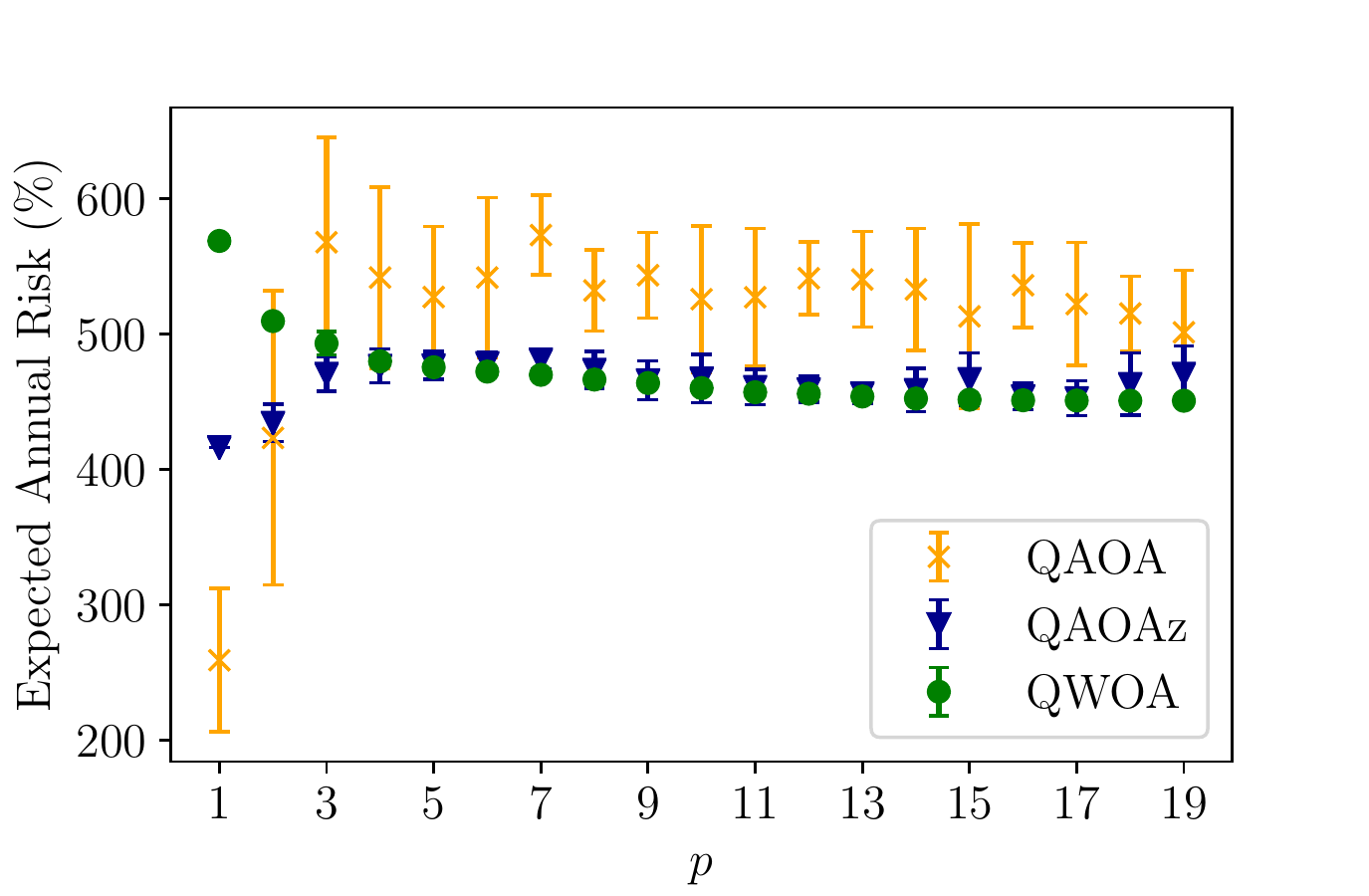}
            \caption{}
            \label{covidrisk}
        \end{subfigure}        
        \begin{subfigure}{.49\textwidth}
            \centering
            \includegraphics[width=\textwidth]{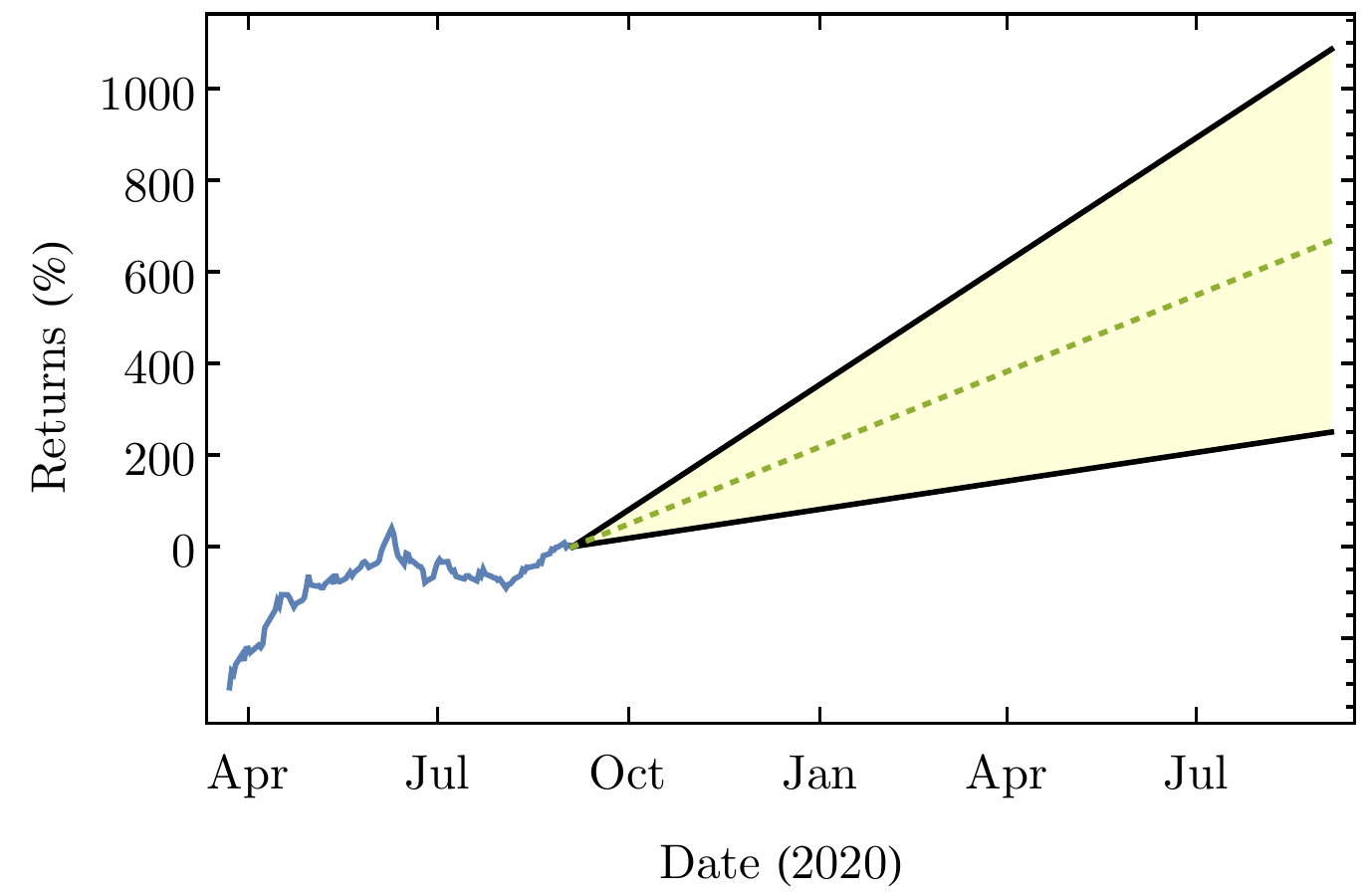} 
            \caption{}
            \label{plot2}
        \end{subfigure}
        \caption{Expected annual portfolio return (a) and expected annual risk (b) for the considered algorithms with increasing $p$ for Data Set B. The historic mean return for Data Set A is shown in (c) along with the projected expected return given by the final QWOA state at $p = 19$; with the yellow region being the 1-$\sigma$ risk region and the dotted line the expected value.}
        \label{fig:covidexpec}
    \end{figure}
    
    \cref{covidret} and \cref{covidrisk} display the expectation value for the returns and expected risk. QWOA yields  the highest expected return for $p >2$. Unlike Data Set A, QAOA is observed to provide the next best expected return for $p > 7$. However, this is associated with a consistently higher expected risk as compared to QAOAz and QWOA, which indicates convergence to a sub-optimal local minima. The mean expected return and risk across across the occupied QAOAz parity bands is $490 \%$ and $542 \%$, taking into account the binomial distribution of the initial state. This is significantly different from the expected risk and return observed for QAOAz in \cref{covidret} and \cref{covidrisk} for $p \geq 3$ which, in combination with the relatively flat QAOAz response to increasing $p$, is indicative of convergence to a state highly influenced by solution degeneracy. As shown in \cref{plot2}, the projected return of the final QWOA state at $p=19$ is a good match for the historic data used, as it again accurately represents the trend and volatility of the data.

    \subsection{Analysis}
    
    {The above results illustrate that the QWOA has an improved rate of convergence (expected solution quality gained per iteration) over the other two approaches. Additionally, the stability of the classical parameter optimisation procedure is higher, as indicated by the smaller error bars. The reasons behind this are multifaceted. The primary improvement is derived from the fact that the combinatorial domain consisting of $\mathcal{M}(n, A)$ valid portfolios is not a binary power in general -- if it were, there would be no `leftover' computational basis states that do not represent a valid solution. In this case, the QWOA would not gain an advantage from considering the valid subspace of solutions. Note that this same advantage holds in general, even under various modifications of the objective function. For example, 
    even if the number of possible asset positions is increased from 3 to 4 such that there is no degeneracy in the qubit encoding, introducing an investment constraint will again reduce the number of valid solutions down from $4^n$ to (in general) a non-binary power. Thus, QWOA will achieve this advantage over any QAOA-based algorithm that cannot restrict to \textit{only} the subspace of non-degenerate valid solutions.}
    
    {The secondary advantage is the choice of the complete graph as the mixer between valid solutions. This leads to an improvement over QAOAz, which has a non solution-transitive connectivity as shown in {\cref{fig:adj-qaoaz}} and does not have a connected solution space. Furthermore, the complete graph is likely the optimal choice of mixer for combinatorial optimisation, based on its proven optimality with respect to spatial database search {\cite{Zalka1999,Childs2004Spatial,Roland2003}}. This holds intuitively, considering the maximal symmetry and connectivity, and the minimal possible graph diameter of 1. These conclusions are supported by the smooth convergence of the QWOA optimisation procedure, where the optimisation of the variational parameters $\vec{t}$ represent optimisation of quantum walk times on the complete graph.}
    
    {Note that the above comments do not depend on portfolio optimisation-specific assumptions. The `energy penalty' constraint encoding approach of the QAOA is applicable to any optimisation problem, and the parity ring mixer formulation of QAOAz is applicable to a wide range of nontrivial combinatorial optimisation domains {\cite{Hadfield_2019}}. Thus, we suggest that these results apply to general `black-box' optimisation problems where the number of valid solutions is not a power of 2. An example of another applicable real-world combinatorial optimisation problem is the well-known Travelling Salesman Problem on $n$ cities {\cite{Applegate2007TSP}}, where the $n!$ feasible solutions necessarily incur leftover states for any qubit solution representation. Rather than introducing degenerate solutions or energy penalties, it is advantageous to restrict strictly to the non-degenerate subspace of valid solutions using the QWOA. As demonstrated by the results in this section, the modest polynomial increase in circuit depth of the QWOA is worth the alleviation of the exponential curse of dimensionality with respect to the classical parameter optimisation.
    
    }
    
    \section{Conclusion}

    This paper carries out a detailed comparison between the performance of the well-known Quantum Approximate Optimisation Algorithm (QAOA), the Quantum Alternating Operator Ansatz (QAOAz), and the newly-developed Quantum Walk Optimisation Algorithm (QWOA) on the NP-hard problem of portfolio optimisation with discrete asset constraints. We perform a detailed analysis of the different mixing operators involved with each technique, and the associated search subspaces. 
    
    Our numerical simulations highlight key advantages of QWOA when compared to both QAOA and QAOAz. QWOA reduces the search space by a significant factor, leading to consistently improved performance in obtaining a high quality portfolio configuration using fewer iterations, and with significantly smaller standard deviation across numerical simulations. Additionally, the global symmetry of the QWOA mixing operator leads to clear advantages in convergence rate and expected solution quality, while QAOA and QAOAz are hindered by bias in the mixing operator over nontrivial feasible solution spaces. {These results show not only the applicability of quantum combinatorial optimisation algorithms to an important real-world problem in the financial realm, but also express the advantages of using the QWOA on arbitrary optimisation problems with complex discrete constraints and associated solution domains.}
    
   \section{Acknowledgements}
    
    This work was supported by resources provided by the Pawsey Supercomputing Centre with funding from the Australian Government and the Government of Western Australia. JBW and SM thank Yuying Li for valuable discussions on computational finance and optimisation. EM and SM acknowledge the support of the Australian Government Research Training Program Scholarship. SM's research was also supported by a Hackett Postgraduate Research Scholarship at the University of Western Australia.
    
	\bibliographystyle{unsrtnat}
	\bibliography{refs}
	


\end{document}